\documentclass[12pt]{article}

\usepackage{amsmath,amssymb,amsthm,amscd,pstricks}
\usepackage{graphicx,tocloft}
\def\figurename{Figure}

\makeatletter

\renewcommand{\fnum@figure}[1]{\textbf{\figurename~\thefigure}:}

\makeatother

\makeatletter

\renewcommand\section{\@startsection{section}{1}{\z@}
                                   {-3.5ex \@plus -1ex \@minus -.2ex}
                                   {2.3ex \@plus .2ex}
                                   {\normalfont\large\bfseries}}
\renewcommand\subsection{\@startsection{subsection}{2}{\z@} 
                                   {-3.25ex\@plus -1ex \@minus -.2ex}
                                   {1.5ex \@plus .2ex}
                                   {\normalfont\normalsize\bfseries}}
\renewcommand\subsubsection{\@startsection{subsubsection}{3}{\z@}
                                   {-3.25ex\@plus -1ex \@minus -.2ex}
                                   {1.5ex \@plus .2ex}
                                   {\normalfont\normalsize\bfseries}}
\renewcommand\paragraph{\@startsection{paragraph}{4}{\z@}
                                   {3.25ex \@plus1ex \@minus.2ex}
                                   {-1em}
                                   {\normalfont\normalsize\bfseries}}

\makeatother

\newdimen\tableauside\tableauside=1.0ex
\newdimen\tableaurule\tableaurule=0.4pt
\newdimen\tableaustep
\def\phantomhrule#1{\haox{\vbox to0pt{\hrule height\tableaurule
width#1\vss}}}
\def\phantomvrule#1{\vbox{\haox to0pt{\vrule width\tableaurule
height#1\hss}}}
\def\sqr{\vbox{%
  \phantomhrule\tableaustep

\haox{\phantomvrule\tableaustep\kern\tableaustep\phantomvrule\tableaustep}%
  \haox{\vbox{\phantomhrule\tableauside}\kern-\tableaurule}}}
\def\squares#1{\haox{\count0=#1\noindent\loop\sqr
  \advance\count0 by-1 \ifnum\count0>0\repeat}}
\def\tableau#1{\vcenter{\offinterlineskip
  \tableaustep=\tableauside\advance\tableaustep by-\tableaurule
  \kern\normallineskip\haox
    {\kern\normallineskip\vbox
      {\gettableau#1 0 }%
     \kern\normallineskip\kern\tableaurule}%
  \kern\normallineskip\kern\tableaurule}}
\def\gettableau#1 {\ifnum#1=0\let\next=\null\else
  \squares{#1}\let\next=\gettableau\fi\next}

\newcommand{\be}{\begin{equation}}
\newcommand{\ee}{\end{equation}}
\newcommand{\bea}{\begin{eqnarray}}
\newcommand{\eea}{\end{eqnarray}}
\newcommand{\ba}{\begin{array}}
\newcommand{\ea}{\end{array}}
\newcommand{\id}{\haox{1\kern-.27em l}}

\newcommand{\RR}{\mathbb{R}}
\newcommand{\half}{ {\textstyle \frac{1}{2}  } }
\newcommand{\al}{\alpha}

\newcommand{\Ga}{\Gamma}
\newcommand{\bet}{\beta}

\newcommand{\de}{\delta}

\newcommand{\ep}{\epsilon}

\newcommand{\si}{\sigma}
\newcommand{\la}{\lambda}

\newcommand{\ze}{\zeta}
\newcommand{\De}{\Delta}
\newcommand{\La}{\Lambda}

\newcommand{\cN}{\mathcal{N}}
\newcommand{\cO}{\mathcal{O}}
\newcommand{\cQ}{\mathcal{Q}}
\newcommand{\cW}{\mathcal{W}}

\newcommand{\cT}{\mathcal{T}}

\newcommand{\cC}{{\mathcal C}}

\newcommand{\D}{{\rm d}}
\newcommand{\pa}{\partial}
\newcommand{\rar}{\rightarrow}

\newcommand{\non}{\nonumber}
\newcommand{\lb}{\langle}
\newcommand{\rb}{\rangle}
\newcommand{\re}{{\rm e}}
\newcommand{\ri}{{\rm i}}
\newcommand{\rd}{{\rm d}}
\newcommand{\ha}{\hat{a}}
\newcommand{\tila}{\tilde{a}}
\newcommand{\SU}{\mathrm{SU}}

\newcommand{\U}{\mathrm{U}}

\newcommand{\ts}{\textstyle}

\begin{document}
\begin{flushright} {CERN-PH-TH/2010-078}\end{flushright}

\begin{center}
\vspace*{3mm}
{\Large\sf
{A \!\&\! B \!\!  model \!\!  approaches \!\!  to \!\!  surface \!\!  operators \!\!  and \!\! Toda  \!\! theories}
}

\vspace*{5mm}
{\large Can Koz\c{c}az${}^\dagger$, Sara Pasquetti${}^\dagger$ and Niclas Wyllard}

\vspace*{5mm}
       ${}^\dagger$ PH-TH division, CERN, CH-1211 Geneva, Switzerland
\vspace*{5mm}

{\tt Can.Kozcaz, Sara.Pasquetti@cern.ch, n.wyllard@gmail.com}

\vspace*{9mm}
{\bf Abstract} 
\end{center}
\vspace*{0mm}
\noindent 
It has recently been argued \cite{Alday:2009b} that the
inclusion of surface operators in  $4d$  $\cN=2$ $\SU(2)$ quiver gauge
theories should correspond to insertions of certain degenerate
operators in the dual Liouville theory. So far only the insertion of a
single surface operator has been treated (in a semi-classical limit).
In this paper we study and generalise this  proposal. Our approach
relies on the use of topological string theory techniques.
On the B-model side we show that the effects of multiple
surface operator insertions in $4d$ $\cN=2$ gauge theories can be
calculated using the B-model topological recursion method, valid beyond the semi-classical limit.
On the mirror A-model side we find by explicit computations that the $5d$ lift of the
$\SU(N)$ gauge theory partition function in  the presence of (one or
many) surface operators is equal to an A-model topological string
partition function with the insertion of (one or many) toric branes. 
This is in agreement with an earlier proposal by Gukov~\cite{Gukov:2009}.
Our A-model results were motivated by and agree with what one obtains by combining the AGT conjecture with the dual interpretation in terms of degenerate operators. 
The topological string theory approach also opens up new
possibilities in the study of  $2d$ Toda field theories.

\vspace{-3mm}

\setcounter{tocdepth}{1}
\tableofcontents

\setcounter{equation}{0}
\section{Introduction}\label{sint}

The new class of 4$d$ $\cN=2$ $\SU(r+1)$ (or $A_{r}$) quiver gauge theories introduced in \cite{Gaiotto:2009} has attracted  much interest. This class of theories can be viewed as arising from the 
$6d$ $A_{r}$  (2,0) theory compactified on $C\times \RR^{1,3}$, where $C$ is a genus $g$ Riemann surface with $n$ punctures, and each puncture is labelled by a Young tableaux with $r+1$ boxes.
The resulting theories, usually denoted $\cT_{(n,g)}(A_{r})$, include not only conventional gauge theories but also more general theories that are not  weakly-coupled gauge theories. 
 
The $4d$ $\cT_{(n,g)}(A_{r})$  theories were subsequently related to $2d$ conformal $A_r$ Toda field theories \cite{Alday:2009,Wyllard:2009} and to  $A_r$ quiver matrix models  \cite{Dijkgraaf:2009}. In particular, the Nekrasov instanton partition functions for the  $4d$ gauge theories are identified with certain conformal blocks in the $2d$ Toda theory \cite{Alday:2009}.
 
Surface operators provide a particularly interesting class of observables in the $\cT_{(n,g)}(A_{r})$ theories. A proposal for how to describe them in the dual $2d$ CFT was presented in \cite{Alday:2009b} (other observables such as Wilson and 't Hooft loops have also been studied \cite{Alday:2009b,Drukker:2009}).  Recently arguments have been presented that indicate that surface operators probe the six-dimensional origin of the theory and may be useful for addressing the problem of classifying conformal $4d$ $\cN=2$ gauge theories \cite{Gaiotto:2009b}. 
In \cite{Alday:2009b} only single surface operator insertions in the $A_1$ quiver gauge theories were treated (and only in a semi-classical limit). In this paper we study and generalise this  proposal. 
 
A central theme in this work is the use of methods and ideas from topological string theory. It is well known that the A-model topological string partition function for the toric Calabi-Yau which engineers a certain $\cN=2$ quiver gauge theory is equal to the Nekrasov instanton partition function for the corresponding (five-dimensional) gauge theory formulated on $\RR^4 {\times} S^1$ \cite{Nekrasov:2002,Iqbal:2003}. The  relation to  topological string theory  offers a powerful computational framework that allow us to use various techniques  (both A and B model) to study surface operators and Toda theories.

On the B-model side we propose and check by explicit compuatations that the effects of multiple surface operator insertions can be calculated using the B-model topological recursion approach \cite{Eynard:2007,Marino:2006,Bouchard:2007}. This method also allows us to go beyond the semi-classical limit in a systematic  order-by-order expansion.  
Furthermore, we argue that summing up all the corrections leads an expression for the partition function in the form of a ``Baker-Akhiezer" function. Compared to the A-model approach, which is  only applicable in a certain patch of the moduli space, the topological recursion method has the advantage that it allows one to explore  the full moduli space of the theory.

On the mirror A-model side we find that when a gauge theory surface operator is present, the A-model topological string partition function should be modified by the insertion of a toric brane. This is in agreement with the earlier proposal by Gukov \cite{Gukov:2009}.
We provide several explicit checks of this proposal. In particular, we show that  open topological string amplitudes with multiple toric branes, computed  using the (refined) topological vertex \cite{Aganagic:2003a, Iqbal:2007}, are related to multiple insertions of degenerate operators in the  $2d$ CFT and thus, according to the proposal of \cite{Alday:2009b}, to multiple insertions of surface operators in the gauge theory. 
 
Our A-model computations were motivated by an argument which, by combining the conjectures in \cite{Alday:2009} and \cite{Alday:2009b}, allows us to obtain a conjectural expression for the Nekrasov instanton partition function for a four-dimensional $\cN=2$ $\SU(N)$ gauge theory in the presence of a surface operator. The resulting expression has the right qualitative features,  involving sums over both conventional four-dimensional instantons as well as ``two-dimensional instantons" and agrees with the result obtained from the A-model calculations with the toric brane insertion.  

The toric point of view can also be used to analyze the $T_N$ ($=\cT_{3,0}(A_{N-1})$) theories. In~\cite{Benini:2009} certain five-brane webs were argued to describe (a five-dimensional version of)  the $T_N$  theories. Viewing these diagrams as toric diagrams, we show explicitly that the topological string partition function for the $T_2$ geometry is a $q$-deformation of the (chiral) Liouville three-point function. The toric point of view may also be used to analyse  three-point functions in the  higher-rank cases, although this is technically much more difficult.
 
The organisation of this paper is as follows. In the next section we briefly review the conjecture in \cite{Alday:2009b} relating surface operators in $4d$ $\cN=2$ $\SU(2)$ gauge theories to insertions of degenerate operators in the dual $2d$ Liouville theory and summarise our conjectures.  
In section \ref{sCFT} we perform some computations in the $A_r$ Toda field theories that will 
be used in later sections, and also make some comments about the extension to degenerate operators  and surface operators in the $\SU(N)$ theories. 

In section \ref{sB} we introduce our approach based on topological recursion in the context of a simple theory known as $T_2$, which has the advantage that explicit calculations are possible. We also discuss a more complicated theory, the $\SU(2)$ theory with $N_f=4$.
 
In section \ref{sA} we study the topological string partition function with toric brane insertions. This corresponds to the five-dimensional version of the instanton partition function with surface operator insertions \cite{Gukov:2009} and generalises the relation between topological string partition functions and Nekrasov instanton partition functions. 
The relation to toric brane insertions provides a constructive method for determining the gauge theory instanton partition function including the effects of surface operators. We illustrate the method explicitly in some examples using the (refined) topological vertex. 

Then in section \ref{sSurf} we use the conjecture in \cite{Alday:2009b} together with the AGT conjecture \cite{Alday:2009} to obtain a closed expression for the Nekrasov partition function in the presence of a surface operator. The resulting expression has the right qualitative features and  appears consistent with the expectation that it should arise from a localisation problem. We also show that it agrees with the result in section \ref{sA} (reduced to four dimensions).

We conclude with a brief summary and outlook. In the appendix some technical details are collected.

\setcounter{equation}{0}
\section{Review of surface operators and summary of conjectures} \label{srev}

Surface operators are objects in $4d$ gauge theories that are supported on $2d$ submanifolds, just like 't Hooft loops are supported on $1d$ submanifolds. Having been largely ignored for a long time, they have recently attracted a bit more interest, see e.g.~\cite{Gukov:2006,Gomis:2007} for some recent work. 

One way to define a surface operator is by specifying the (singular) behaviour of the fields in the gauge theory near the submanifold where the surface operator is localised.  In general, there are several different consistent possibilities leading to different surface operators.  It has been argued that there is a correspondence between different surface operators localised on an $\RR^2$ submanifold and the so-called Levi subgroups of the gauge group \cite{Gukov:2006}.  In the case of $\cN=2$ gauge theories, the surface operator depends on one complex parameter for each $\U(1)$ factor in the Levi subgroup (see \cite{Gukov:2006,Alday:2009b,Gaiotto:2009b} for further details). In this paper we focus on  the simplest type of surface operator which depends on only  one parameter.

Given the AGT relation between the $4d$  $\cT_{(n,g)}(A_{r})$ theories and the $2d$ $A_r$ conformal Toda field theories, a natural question to ask is what the surface operators correspond to in the $2d$ CFT. This question was addressed in \cite{Alday:2009b} where it was argued that inserting surface operators in $\SU(2)$ quiver gauge theories should correspond to inserting vertex operators corresponding to certain degenerate operators/null states into the relevant correlation function in the dual Liouville theory. 

Assuming this correspondence, the authors of \cite{Alday:2009b} then went on to argue how,  in a certain semi-classical limit, the surface operators should affect the partition function in the gauge theory.  Let us briefly recall the line of reasoning here. 
In the absence of surface operators  the AGT relation \cite{Alday:2009} equates  (up to an overall factor)  the Nekrasov partition function to a certain conformal block in the $2d$ Toda theory.  The momenta of the Toda theory primary fields, $\al_i$, are related to the masses,  $m_i$, in the gauge theory (the exact form of the relations depend on conventions for the gauge theory masses). Furthermore, the internal momenta in the chiral block, denoted by $\si_k$ in this paper, are linearly related to the $a_k$ Coulomb moduli. Finally, the parameters $\ep_1$ and $\ep_2$ in the instanton partition function are related to the parameter $b$ in the Toda theory via
\be
b = \ep_1 \,,\qquad \frac{1}{b} = \ep_2 \,.
\ee 

For simplicity, let us focus on the $\mathcal{N}=2$ $\SU(2)$ theory with $N_f=4$. This theory is a $\cT_{4,0}(A_1)$ theory that can be obtained by compactifying the $6d$,  $(2,0)$ supersymmetric theory  on $C$, which in this case is a sphere with four punctures.  
The low energy dynamics of the gauge theory is encoded in the Seiberg-Witten curve $\Sigma$ (a double cover of~$C$) equipped with 
the Seiberg-Witten differential $\lambda_{SW}$. The  Seiberg-Witten curve can be written~\cite{Gaiotto:2009} as:
\be \label{SWcurveSU2}
x^2 + \psi_2(z)=0\,,
\ee
and the Seiberg-Witten differential is $\la_{SW}=x \, \D z$. 
The  Nekrasov instanton partition function of the gauge theory has the general form:  
\be \label{sublead}
Z  = \re^{-\frac{1}{\epsilon_1 \epsilon_2} \left(F_0 + (\epsilon_1+\epsilon_2) H_{1/2}
+ (\epsilon_1+\epsilon_2)^2 H_1  + \epsilon_1\epsilon_2 F_1 +\cdots \right)}.
\ee
Each term in this expansion is identified, via the AGT relations,  
with a corresponding term in the 
semi-classical expansion of  the four-point conformal block in the Liouville theory on $C$.
The (conformal) Liouville theory  has the central charge
$c = 1 + 6 \,\mathcal{Q}^2$ where $\cQ=b+1/b$, and a set of primary fields,  $V_{\al} = \re^{2 \al \phi}$, with
conformal dimensions $\De(\al)= \al ( \cQ-\al)$, i.e.~$L_0 V_\al =
\De(\al) V_{\al}$. 

  It is convenient to fix three of the points to $0, 1,
\infty$ and use  a bra--ket notation which has the property $\lb
\al|\al \rb=1$, and is such that
\be \label{4ptbraket}
\lb \al_1| V_{\al_2}(1) V_{\al_3}(\ze) |\al_4 \rb = \lb V_{Q-\al_1}(0)
V_{\al_2}(1) V_{\al_3}(\ze) V_{\al_4}(\infty) \rb\,.
\ee
The semi-classical expansion is obtained on the CFT side 
 by scaling all momenta\footnote{An equivalent alternative point of view, used in \cite{Alday:2009b}, is to rescale the $\ep_i$ appearing in the gauge theory partition function as $\ep_i\rar \hbar \, \ep_i$ leaving the momenta unchanged.}  (both internal and external) as $\al_i \rar \al_i/\hbar$ and  performing a double expansion  in $\hbar$ and $  \mathcal{Q}=b+\frac{1}{b}$,
related  to the {\it topological} (or genus) expansion of the dual gauge theory and to its $\mathcal{Q}$uantisation,  respectively. The conformal block (suitably normalised) corresponding to (\ref{4ptbraket}) takes the form
\be \label{CFT4pt}
\re^{ -\frac{1}{\hbar^2}F_0  - \frac{\mathcal{Q}}{\hbar} H_{1/2} - F_1 - \mathcal{Q}^2 H_1+\cdots  }.
\ee

In \cite{Alday:2009b}, it was argued that in the presence of a surface operator and in the semi-classical limit, the corresponding CFT expression should behave as
\be \label{Znull1}
Z_\mathrm{null}= \frac{ \langle \al_1 | V_{\al_2}(1) V_{\al_3}(\ze) V_{-b/2}(z)   | \al_4  {+} \frac{b}{2} \rangle }{  \langle V_{\al_1} |  V_{\al_2}(1) V_{\al_3}(\ze) | \al_4   \rangle }
= \re^{- \frac{b}{\hbar}G_{0}(z)  +\cdots}\,.
\ee
Here we have normalised the expression with respect to the chiral four-point function, (\ref{4ptbraket}) (if this is not done there will in general be additional terms at order $\hbar^{-1}$ coming from the expansion (\ref{CFT4pt}) which do not go to zero as $b\rar 0$). We have also shifted $\al_4$ in the numerator by $b/2$. This shift does not affect the $G_{0}(z)$ term, but is important at higher orders. The location of the degenerate operator insertion, $z$, is related to the parameter describing the surface operator \cite{Alday:2009b}.

In the Liouville theory, the primary field $V_{-b/2}$ satisfies the null state condition $(L_{-1}^2+b^2L_{-2})V_{-b/2}=0$. This implies that 
\be
\pa_z ^2 \langle \al_1 | V_{\al_2}(1)V_{\al_3}(\ze)  V_{-b/2}(z) | \al_4   \rangle   
+ b^2  \langle \al_1 | V_{\al_2}(1)  V_{\al_3}(\ze)  T(z) V_{-b/2}(z)| \al_4   \rangle  =0\,.
\ee
It was shown in \cite{Alday:2009} that
\be
\label{cftdiff}
 \langle \al_1 | V_{\al_2}(1)V_{\al_3}(\ze) T(z) V_{-b/2}(z)  | \al_4   \rangle \rar  
 \frac{\psi_2(z)}{\hbar^2} \langle \al_1 | V_{\al_2}(1)  V_{\al_3}(\ze) V_{-b/2}(z) | \al_4   \rangle \,,
 \ee 
 in the semi-classical limit $\hbar \rar 0$. Note that $\psi_2(z)$ appearing here is the same as for the theory without the surface operator insertion (this follows from the limit we are taking).
 
Furthermore, a WKB-type argument implies that, to leading order, acting with $\pa_z^2$ on (\ref{Znull1}) brings down $\frac{b^2}{\hbar^2} (\pa_z G_{0}(z))^2$. Collecting the above facts we obtain
\be 
(\pa_zG_{0})^2+ \psi_2(z)=0\,.
\ee

 By comparing this expression to the Seiberg-Witten curve, (\ref{SWcurveSU2}),
  one finds that $\pa_zG_{0}(z)\D z$ is equal to the Seiberg-Witten differential,  $\lambda_{SW}$, on one of the two sheets of the Seiberg-Witten  curve. By integration one finally finds \cite{Alday:2009b}
\be \label{Gz}
G_{0}(z) = \int^z \! x(z') \, \D z'\,.
\ee 
This argument shows that (in the semi-classical limit) the effect of the surface operator insertion is contained in the function $G_{0}(z)$ which can be determined in terms of the Seiberg-Witten data (i.e.~the curve and the differential) via the above formula.

We now make the observation that (\ref{Gz}) is nothing but the B-model topological string disk amplitude\footnote{This observation is implicit in \cite{Alday:2009b}.  }.  

Let us elaborate  on this point. Consider the type II string theory setup where the  four-dimensional $\cN=2$ gauge theory is engineered by compactification on a  toric  Calabi-Yau three-fold.
On the  type IIA side it is well known \cite{Iqbal:2007,Awata:2005,Taki:2007} that the topological string partition function computed using the refined topological vertex reproduces the Nekrasov    instanton partition function in five dimensions (and in the field theory limit, the four-dimensional partition function).

On the type IIB side the local Calabi-Yau geometries are mirror manifolds of toric three-folds  of the form:
\be
Y_u:\qquad w w'= H(x,y;u),
\ee
where $H(x,y;u)=0$ is the {\it mirror curve}, a family of algebraic curves  parameterized by the complex structure parameter  $u$, embedded in $\mathbb{C}^*\times \mathbb{C}^*$. 
Open topological B-model amplitudes have boundary conditions provided by the 
mirror of  toric  A-branes  that  wrap  holomorphic curves in $Y_u$ with trivial bundles, defined by:
\be
w' = 0 = H(x_0,y_0;z).
\ee
The open moduli space corresponds to deformations of the B-brane in $Y_u$
 which are parameterized by the points $(x_0,y_0) \in H(x,y,u)$. 
As a result, the moduli space of the open B-model coincides with the mirror curve.

The disk amplitude \cite{AV,AKV} is obtained from the line integral of the one--form on the mirror curve, obtained by solving $y$ as a function of $x$ in $H(x,y;u)=0$:
\be
A^{(0)}_1(z)=\int^z \log(y(x;u)) \frac{\D x}{x}.
\ee
This is exactly the way the surface operator is evaluated, in the 
semi--classical limit, in the gauge theory, with the difference that 
(\ref{Gz}) involves the {\em M-theory}  differential $\lambda_{SW}$
 rather than the {\it engineering} one (the field theory limit of $\log(y(x;u)) \frac{\D x}{x}$).
 We will revisit this point in section \ref{torsurf}.

We now summarise our results and conjectures. As in \cite{Gukov:2009} we argue  that a surface operator in the gauge theory is realized in the topological string setting by the insertion of a {\em single} toric A brane. 
We gather evidence in support of this conjecture in section \ref{sA} based on explicit topological vertex computations. 
In particular, we show that the topological string partition function obtained using the  {\it refined} vertex \cite{Iqbal:2007} agrees, in the four-dimensional limit, with the expression for a normalised  CFT correlation function with degenerate operator insertions, while the computation using the usual vertex \cite{Aganagic:2003a} captures the same expression when $\mathcal{Q}=0$. 

The relation to toric branes is also natural from another viewpoint. In~\cite{Schiappa:2009} it was shown that, using the proposal in \cite{Dijkgraaf:2009}, degenerate state insertions in the Liouville CFT correspond, in the matrix model language, to insertions of $\det(z-\Phi)=e^{\varphi(z)}$ where $\varphi(z)$ is a chiral scalar. Insertions of a similar type were argued to be related to toric branes in \cite{Aganagic:2003a,Aganagic:2003b}. 

The topological string realisation of the AGT conjecture with surface operators  makes  the generalization of the original proposal to multiple insertions immediate:  multiple surface operator insertions correspond to multiple toric brane insertions.

On the B-model side open amplitudes can be computed using the {\it topological recursion} method \cite{Marino:2006,Bouchard:2007}. This method extends the original result (\ref{Gz}) 
to the case with multiple insertions of surface operators as well as beyond
the semi-classical limit.

Compared to the A-model (topological vertex) approach, the B-model  
(remodeling) approach has the disadvantage of being perturbative in $\hbar$.
However, it has the great advantage of being non-perturbative in the complex structure parameters. As explained in \cite{ABK,Bouchard:2007} this allows one to study the duality frame transformations of the amplitudes. In the AGT context this is particularly interesting because of the relation to the
 S-duality transformations in the gauge theory. 
In particular, the B-model setup makes feasible the computation of amplitudes 
in patches corresponding to strong coupling limits of the gauge theory, where 
the A-model setup, and consequently the Nekrasov partition function, cannot be used.

\setcounter{equation}{0} 
\section{CFT approach: Degenerate operators  in Toda field theory} \label{sCFT}

In this section we derive some results for  $A_r$ conformal Toda field theories that will be useful in later sections. We first study the structure of the correlation functions in the Liouville theory  with multiple  insertions of degenerate operators, and work out the details for the two specific examples which correspond to the $\cT_{3,0}(A_1)$ and  $\cT_{4,0}(A_1)$ gauge theories. Then we briefly discuss degenerate operators in the higher rank Toda theories, using the $A_2$ theory as an example. These degenerate operators are relevant for surface operator insertions in $\SU(N)$ gauge theories.  Finally, we discuss some facts about the $\cT_{3,0}(A_r)$ theories and comment on the number of parameters in the $A_r$ Toda theory (chiral) three-point functions. 

\subsection{The $\cT_{3,0}(A_1)$ theory} \label{sill}

In this section we discuss the $\cT_{3,0}(A_1)$ (a.k.a.~$T_2$) theory. In the language of \cite{Gaiotto:2009} this theory arises from three punctures on a sphere and is simply a free theory with four hypermultiplets. Although very simple it is still non-trivial enough to allow us to illustrate our approach and methods. A bonus is that we will be able to perform exact calculations. Later on we study more complicated examples. 

The Seiberg-Witten curve of the $T_2$ theory is $x^2 + \psi_2(z)=0$, where (see e.g.~\cite{Schiappa:2009})
\be \label{T2SW}
 \psi_2(z)= \frac{ -\al^2_1 z(z-1) - \al^2_2 z - \al_3^2 (1-z)  }{z^2(1-z)^2}\,.
\ee
Another way to obtain this expression is from the Liouville theory via \cite{Alday:2009}
\bea
\label{qc}
 \hat\psi_2(z) &\equiv& \frac{\lb \al_1|T(z)V_{\al_2}(1)|\al_3\rb}{\lb \al_1|V_{\al_2}(1)|\al_3\rb} = \sum_{n=0}^{\infty} z^{-n-2} \frac{\lb \al_1|L_n V_{\al_2}(1)|\al_3\rb}{\lb \al_1|V_{\al_2}(1)|\al_3\rb}  \non \\
&= & \frac{\De(\al_1)z(z-1)+\De(\al_2)z+\De(\al_3)(1-z) }{z^{2}(z-1)^2}\rar  \frac{\psi_2(z)}{\hbar^2}\,,
\eea
where in the last step we rescaled $\al_i \rar \al_i/\hbar$ and took the semi-classical limit $\hbar\rar0$. 
We will comment later about the role of $\hat\psi_2(z)$ and the `quantum'  curve.  

The partition function for the $T_2$ theory does not have any gauge theory instanton corrections.
However, we can still add surface operators to the theory (at least formally), which will lead to non-trivial corrections. 
In the simplest example one adds a single surface operator, which using the arguments in \cite{Alday:2009b}, leads to the four--point function in the Liouville theory where one of the $\al_i$ is equal to $-b/2$. More precisely, the expression we are interested in is:
\be \label{bVVV}
Z_{\rm null}(z)= \frac{\lb \al_1 |  V_{\al_{2}} (1)  V_{-b/2}(z)| \al_3{+}\frac{b}{2}  \rb}{ \lb \al_1 |  V_{\al_2} (1) | \al_3  \rb}   \,.
\ee
As already mentioned in the previous section, the shift of $\al_3$ in the numerator is important and ensures a simple behaviour when $z\rar0$. 

It is well known that in the Liouville theory the four--point function with one degenerate insertion 
satisfies the hypergeometric differential equation; this result implies that
\be \label{2F1}
Z_{\rm null}(z)=z^{b\al_3}(1-z)^{b\al_2} \,{}_2 F_1(A_1,A_2;B_1;z)\,,
\ee
where 
\be \label{ABC2}
A_1 = b\, (\al_1+\al_2-\al_3)\,,\quad  A_2=b\, (\al_1+\al_2+\al_3 - \cQ )\,, \quad B_1= 2\,  b \, \al_1  \,.
\ee
We now introduce the following semiclassical expansion 
\be
\label{lt21}
Z_{\rm null}(z)=\exp\left\{\frac{b}{\hbar}G_{0}(z)+ b^2 G_{1}(z)+ b^3 \hbar \, G_{2}(z)+\mathcal{O}(\hbar^2)\right\}
\ee
with
\be
\label{gqexp}
G_{i}(z)=\sum_{n\ge 0}G_{i}^n(z)\mathcal{Q}^n.
\ee
Since $Z_{\rm null}(z)$ is known exactly in this case, (\ref{2F1}), it is straightforward to obtain expressions for the $G_i(z)$'s by rescaling $\al_i\rar \al_i/\hbar$ and Taylor expanding. For instance,
\be \label{g01}
G^0_0(z) =  -\frac{(\al_3^2 {-} \al_1^2 {-} \al_2^2)}{2 \al_1} z - \frac{(\al_3^4 + 2\, \al_3^2 \al_1^2 - 3\, \al_1^4 - 2\, \al_3^2 \al_2^2 -   6\, \al_1^2 \al_2^2 {+} \al_2^4)}{16 \al_1^3} z^2 + \ldots
\ee

Next we turn to the cases corresponding to $k>1$ insertions of $V_{-b/2}$ i.e.
\be \label{LNpt}
Z_{\rm null}(z_1,\cdots,z_k) = \frac{ \lb \al_1|  V_{\al_2} (1) V_{-b/2}(z_1) \cdots V_{-b/2}(z_k)  |\al_3 {+}k \frac{b}{2} \rb }{\lb \al_1|  V_{\al_2} (1) |\al_3 \rb} \,.
\ee
This expression can also be calculated exactly. The numerator was determined in \cite{Schiappa:2009} using the connection to matrix models together with an earlier result of Kaneko \cite{Kaneko:1993}. Using this result we find
\be \label{genhyp}
Z_{\rm null}(z_1,\cdots,z_k) = \left( \prod_{i=1}^k z_i^{b\al_1}(1-z_i)^{b\al_2}  \right) {}_{2}F^\bet_1(A_1,A_2;B_1;z_1,\ldots,z_k)\,.
\ee
where 
\be \label{ABCgen}
A_1 = b\,(\al_1+\al_2-\al_3)\,,\quad  A_2=b\,(\al_1+\al_2+\al_3 -\cQ )\,, \quad B_1= 2\, b \, \al_1  \,,
\ee
and the function ${}_{2}F^\bet_1(A_1,A_2;B_1;z_1,\ldots,z_k)$ is  the generalised hypergeometric function defined as \cite{Kaneko:1993}
\be \label{hyperJack}
{}_{2}F^\bet_1(A_1,A_2;B_1;z_1,\ldots,z_k) 
= \sum_{\xi}  \frac{[A_1]^{\bet}_\xi[A_2]^{\bet}_\xi}{[B_1]^{\bet}_\xi} \frac{\cC^{\bet}_\xi(z_1,\ldots,z_k)}{|\xi|!}\,,
\ee
where the sum is over all partitions $\xi= \{ \xi_{i} \}$ with at most $k$ parts,
\be
[X]_\xi^\bet = \prod_i (X-\frac{1}{\bet}(i-1))_{\xi_i}\,,
\ee
and $\cC_\xi^{\beta}(z_1,\ldots,z_k)$ is a  Jack polynomial with a particular normalisation\footnote{Jack polynomials also play an important role in the recent matrix model developments \cite{Itoyama:2010}.}. For the convenience of the reader we list the first few Jack polynomials in appendix \ref{jack}.

The semi-classical expansion of these more general expressions can also be obtained. In particular, for two insertions one has:
\be
\label{lt22}
Z_{\rm null}(z_1,z_2)=\exp\Big\{ \frac{b(G_{0}(z_1)+G_{0}(z_2))}{\hbar}+
b^2 G_1(z_1,z_2)
 + \cO(\hbar) \Big\}.
\ee
Notice that in addition, each term has a  $\mathcal{Q}$ expansion:
 \be
 G_{i}(z_1,\cdots,z_k)=\sum_{n\ge 0}\mathcal{Q}^n G^{n}_i(z_1,\cdots,z_k).
 \ee
As an example, by Taylor expansion we find
\be \label{g02}
G^0_1(z_1,z_2) = 
-\frac{ (\al_1 {-} \al_2 {+} \al_3) (\al_3 {-} \al_1 {+} \al_2) (\al_1 {+} \al_2 {-} \al_3) (\al_3 {+} \al_1 {+}   \al_2)}{16 \al_1^4 }(\frac{z_1^2}{2} {+} \frac{z_2^2}{2} {+} \,  z_1 z_2 ) +...
\ee
Similarly, for three insertions one has:
\bea
\label{lt23}
&&Z_{\rm null}(z_1,z_2,z_3)=\exp\Big\{ \frac{ b\left(G_{0}(z_1)+G_{0}(z_2)+
G_{0}(z_3)\right)}{\hbar} \\&&+b^2\left(  G_1(z_1,z_2)
+ G_1(z_1,z_3)+ G_1(z_2,z_3)\right)+b^3\hbar G_2(z_1,z_2,z_3)
 + \cO(\hbar^2) \Big\}. \non
\eea

\subsection{The $\cT_{4,0}(A_1)$ theory} \label{sill2}

As is well known, higher--point correlation functions in the Liouville theory can be related
to the three--point functions of primary fields, which therefore
determine the entire theory \cite{Belavin:1984}.  
For instance, inserting a complete set of states into the (chiral) four-point function one finds
\be \label{cb}
\lb \al_1| V_{\al_2}(1) V_{\al_3}(\ze) |\al_4 \rb = \int \D \si
\sum_{{\bf n},{\bf n}'}
   \lb \al_1| V_{\al_2}(1) |{\bf n};\si \rb X^{-1}_{ \bf n; \bf
n'}(\si)  \lb {\bf n}';\si| V_{\al_3}(\ze) |\al_4 \rb  \,.
\ee
where $X_{ \bf n; \bf n'}(\si)=    \lb \bf n; \si |\bf n' ; \si \rb $
and the intermediate states\footnote{Throughout this paper we will
label the internal momenta by $\si$ reserving the symbol $\al$ for the
external momenta.} $|{\bf n}; \si\rb$ are descendants of the primary
state labelled by $\si$ i.e.~
\begin{eqnarray} \label{dessta}
|\si,{\bf n}\,\rangle&\equiv& L_{-n_{1}}L_{-n_{2}}\mathellipsis
L_{-n_{r}}|\si\rangle\,,\qquad (1\leq n_{1}\leq
n_{2}\leq\mathellipsis\leq n_{r}) \\
\langle\si,{\bf n}\,|&\equiv&\langle\si|
L_{n_{s}}L_{n_{s-1}}\mathellipsis L_{n_{1}}\,, \quad \qquad (1\leq
n_{1}\leq n_{2}\leq\mathellipsis\leq n_{s}) \non
\end{eqnarray}
where ${\bf n}=(n_1,n_2,\ldots )$ and $|{\bf n}|=n_{1}+n_{2}+\mathellipsis$.
It can be shown that $\lb {\bf n}',\si| V_{\al_3}(\ze) |\al_4 \rb$ is
proportional to $\ze^{|{\bf n'}|}\lb \si| V_{\al_3}(\ze) | \al_4 \rb$
\cite{Belavin:1984};  hence (\ref{cb}) can be calculated
perturbatively.

For later purposes, we are interested in computing chiral $k+4$-point
functions with $k$ insertions of the degenerate operator 
$V_{-\frac{b}{2}} $. To this end it turns out to be convenient to make
use of the fact that we already know how to compute chiral $k+3$-point
functions with $k$ degenerate operator insertions. Thus we only
insert a single complete set of states and consider
\bea \label{kconfb}
&&\!\! \!\!  Z_{\mathrm{null}}(\ze,z_1,\ldots,z_k)= \non \\
&&\!\! \!\! 
\frac{ \sum_{{\bf n},{\bf n}'}  \lb \al_1| V_{\al_2}(1) V_{-\frac{b}{2} }(z_1) \cdots
V_{-\frac{b}{2} }(z_k)  |{\bf n};\si+\frac{kb}{2} \rb X^{-1}_{ \bf n; \bf n'}(\si)
\lb {\bf n}';\si| V_{\al_3}(\zeta) |\al_4 \rb }{  \sum_{{\bf m},{\bf m}'}  \lb \al_1|
V_{\al_2}(1) |  {\bf m};\si \rb X^{-1}_{ \bf m; \bf m'}(\si)
\lb {\bf m}';\si|  V_{\al_3}(\ze) |\al_4 \rb  }  \,. 
\eea
This construction leads to a result   symmetric in the $z_i$,
which  is important for the dual gauge theory and topological string
interpretations that will be discussed in later sections, where this
symmetry is manifest.

A few comments about (\ref{kconfb}) are in order. As in previous sections, we have shifted the momentum of the state closest to the degenerate operator insertions. However, in this case there may be additional  modifications needed, since the state now also involves descendants.  
We have chosen to insert the degenerate operators to the left of $ V_{\al_3}(\ze) $. Another possibility would have been to insert them to the right. This possibility is discussed in later sections. 

To compute the expression
\be
 \lb \al_1| V_{\al_2}(1) V_{-\frac{b}{2} }(z_1) \cdots
V_{-\frac{b}{2} }(z_k)  |{\bf n};\si \rb \,,
\ee
one makes use of the definition of  $|{\bf n};\si \rb$ (\ref{dessta})
and moves the $L_{-n_i}$'s to the left using the relations:
\bea \label{LV}
[L_n, V_{\al}(z)] &=&  z^n[ \,(n+1) \, \De(\al) \, V_\al(z) + z\,
(\pa_z V_\al(z)) \,] \non\\
&=&  z^n( \,n\, \De(\al)\,  V_\al(z) +  [L_0,V_\al(z)]  \,) \,,
\eea
The first relation is used until one reaches $ V_{\al_2}(1)$ at which
point the second relation is used. One then moves the remaining $L_0$
to the right, again using the first relation. As an example, to linear
order in $\ze$ this procedure gives
\bea
&&\!\!\! \!\!\!  \zeta \, (\De(\si)+\De(\al_2)-\De(\al_1) ) \times \\
&&\!\!\! \!\!\!  \bigg[ \sum_{i=1}^k (z_i{-}1) \pa_{z_i} {+} k
\De(-b/2) {+} \De(\si) {+} \De(\al_3){-}\De(\al_4) \bigg]  \lb \al_1|
V_{\al_2}(1) V_{-\frac{b}{2} }(z_1) \cdots  V_{-\frac{b}{2} }(z_k)  |
\si \rb \non
\eea
and since
\be
\lb \al_1| V_{\al_2}(1) V_{-\frac{b}{2} }(z_1) \cdots  V_{-\frac{b}{2}
}(z_k)  | \si \rb
\ee
is known exactly, cf.~(\ref{genhyp}), one can plug in this result and carry out
the differentiations.  The result of this procedure is exact in the
$z_i$, but perturbative in $\ze$.

As above, we can define the $G_i(\ze,z_1,\ldots,z_k)$'s. As an example, we write out the first few terms in the semi-classical expansion of $G_1^0(\ze,z_1,z_2)$ when $k=2$ explicitly:
\bea 
\label{g02su2}
&&G^0_1(\ze,z_1,z_2) = -\frac{(\al_3 {+} \al_4 {-} \si) (\al_3 {-} \al_4 {+} \si) (\al_4 {-} \al_3 {+} \si) (\al_3 {+} \al_4 {+} \si)}{
 16 \si^4} z_1 z_2 \\&&
 +\, \ze \, \frac{(\al_3^2 - \al_4^2) (\al_3^4 {-} 2 \al_3^2 \al_4^2 {+} \al_4^4 {-} 2  \si^2 \al_3^2 {-}2 \si^2  \al_4^2 {+} \si^4) (-\al_1^2 + \al_2^2 + \si^2)}{16 \si^8} z_1 z_2 + \cO(\ze^2) \non 
 \eea
Here we have only written terms involving both $z_1$ and $z_2$; terms not of this type are sensitive to the issue discussed below (\ref{kconfb}), and possibly have other ambiguities as well, and will not be considered in this paper.

\subsection{ $\cW$-algebra degenerate states and $\SU(N)$ surface operators }\label{sW}

In this section we briefly discuss the extension to the higher rank case, i.e.~surface operators in $\SU(N)$ quiver gauge theories. 
The $\SU(N)$ quiver gauge theories are dual to the $A_{N-1}$ Toda field theories \cite{Alday:2009,Wyllard:2009}. In the $A_r$ Toda field theory the symmetry algebra is the $\cW_{r+1}$ algebra with central charge $c=r +12 (b+\frac{1}{b})^2 \lb \rho,\rho \rb$, where $\rho$ is the Weyl vector. In the particular case of the $A_2$ theory the $\cW_3$ algebra has generators $\cW_3(z)\equiv \cW(z)$ and $\cW_2(z)\equiv T(z)$ with mode expansions
\be
 T(z) = \sum z^{-n-2} L_{n} \,, \qquad \cW(z) = \sum z^{-n-3} W_{n}  \,,
\ee
and the central charge is $c= 50 + 24 b^2 + 24\frac{1}{b^2}$. 
It is known that in this theory $V_{\al}$ with $\al = -b \La_1$, where $\La_1$ is the weight of the fundamental representation, is a degenerate operator. This degenerate operator satisfies \cite{Bajnok:1992,Fateev:2007b}
\bea \label{W3null}
&&\!\!\!\! \left( W_{-1} - \frac{3w}{2\De} L_{-1} \right) V_\al =0\,, \non \\
&&\!\!\!\!  \left( W_{-2} - \frac{12w}{\De(5\De+1)} L_{-1}^2  + \frac{6w(\De+1)}{\De(5\De+1)} L_{-2} \right) V_\al =0 \,, \\
&&\!\!\!\!  \left( W_{-3} - \frac{16w}{\De(\De{+}1)(5\De{+}1)} L_{-1}^3  + \frac{12w}{\De(5\De{+}1)} L_{-1} L_{-2} + \frac{3w(\De{-}3)}{2\De(5\De+{1})} L_{-3} \ \right) V_\al =0 \,,\non
\eea
where
\be
\De = -(1+\frac{4b^2}{3}) \,,  \qquad w= -i \frac{(3+4b^2)}{9} \sqrt{\frac{2(3+5b^2)}{3(5+3b^2)} } \,.
\ee
To proceed we need the general results 
\be
(L_{-1}V) = \pa \,V \,, \qquad (L_{-n-2}V) = \frac{1}{n!}(\pa^nT \,V) \,, \qquad (W_{-n-3}V) = \frac{1}{n!}(\pa^n \cW \,V) \,.
\ee 
The conventionally normalised primary field $\cW(z)$ is related to the fields $U_3(z)$ and $U_2(z)\equiv T(z)$ appearing in the Miura transform as (see e.g.~\cite{Bouwknegt:1992})
\be
\cW = i \sqrt{\frac{48}{22 +5c}} \left( U_3(z) - \frac{1}{2} \pa U_2 \right) .
\ee
It was shown in \cite{Kanno:2009} that in the semiclassical limit
\be
\lb U_n(z) V \cdots V \rb \rar  \frac{\psi_n(z)}{\hbar^2} \lb V \cdots V \rb \,.
\ee
Here $\psi_n(z)$ are the objects appearing in the Seiberg-Witten curve \cite{Gaiotto:2009}
\be \label{SUNSW}
x^{r+1} + \psi_ 2(z) x^{r-1} + \cdots + \psi_{r+1}(z) = 0 \,.
\ee
In order to be consistent with the semi-classical approximation we need to neglect $\lb \pa T(z) V \cdots V \rb $ compared to  $\lb U_3(z) V \cdots V \rb $. This means that we may neglect the $L_{-3}$ term in the last equation in (\ref{W3null}) and also  implies that $L_{-1}L_{-2}$ effectively can be replaced by $L_{-2}L_{-1}$. Considering a correlation function with a degenerate state insertion and using the same argumentation as in the rank one (Liouville) case we then find (in the semi-classical limit)
\be \label{GSW}
(G_{0}'(z))^3 + G_{0}'(z) \psi_2(z) + \psi_3(z) =0\,.
\ee
Comparing this result to the Seiberg-Witten curve (\ref{SUNSW}) with $r=2$ we find that $G_{0}'(z)$ is equal to the Seiberg-Witten differential on one of the three sheets of the Seiberg-Witten curve. Thus the expression (\ref{Gz}) appears to be universal (this was also argued in \cite{Alday:2009b} from an M-theory perspective).

Note that the limit we are taking is different from the limit sometimes used in Toda theories (see e.g.~\cite{Fateev:2007b}), which corresponds to rescaling the momenta as  $\al \rar \al/b$ and taking $b \rar 0$. In this limit an equation very similar to the above result (\ref{GSW})  was derived in~\cite{Fateev:2007b}, see eq.~(3.21). The difference stems from the way the limits were taken. In~\cite{Fateev:2007b} $\lb \pa T(z) V \cdots V \rb $ was not neglected compared to  $\lb U_3(z) V \cdots V \rb $. 

\subsection{Number of parameters in the general Toda three-point function}  \label{Toda3}

The $T_N$ (or $\cT_{3,0}(A_{N-1})$) theory should presumably correspond to some (chiral) three-point function in the $A_{N-1}$ Toda theory. It is known (see e.g.~\cite{Bowcock:1993,Fateev:2007b,Wyllard:2009}) that, except for the rank one case, the three-point functions of $\cW$ primary fields do not determine the higher-point correlation functions. Instead further data is required. For instance, in the rank two case (corresponding to the $\cW_3$ algebra), one also needs e.g.~the additional three-point functions (where $n$ is a positive integer)
 \be \label{W1n}
\langle \al_1 | (W_{-1}^n V_{\al_2}) | \al_3 \rangle \,.
 \ee
One way to understand this fact is to note that the number of Ward identities is $8$ (corresponding to $L_0$, $L_{\pm 1}$, $W_0$, $W_{\pm 1}$ and $W_{\pm 2}$) whereas the number of ``basis states" is 9 (corresponding to $L^n_{-1}V_{\al}$, $W^n_{-1}V_{\al}$, and $W^n_{-2}V_{\al}$ for each of the three $\al_i$'s). Therefore one of the basis states is left unconstrained and can be chosen e.g.~as in (\ref{W1n}) above.

There is a slight puzzle here since the $T_3$ theory depends on seven (continuous) parameters (six masses and one Coulomb modulus), whereas the set of three-point functions depend on six parameters coming from the $\al_i$ plus the set of positive integers labelling the expressions (\ref{W1n}). However, by Taylor expansion, a function of one variable whose dependence is analytic contains the same amount of information as a set of parameters labelled by a positive integer. Therefore, it seems natural to  package the set of three-point functions (\ref{W1n}) into a generating function. 

For a general $N$ the corresponding counting works as follows. The remaining number of unconstrained sets of positive integers (number of analytic parameters) is
\be
3 \times \sum_{i=1}^{N-1} i - \sum_{i=1}^{N-1} (2 i+1)  = \frac{(N-1)(N-2)}{2} \,,
\ee
which together with the number of parameters, $3(N-1)$, in the three $\al_i$ gives the total number of parameters
\be
\frac{(N+4)(N-1)}{2} \,.
\ee
This number precisely agrees with the number of parameters in the $T_N$ theory, as extracted from \cite{Gaiotto:2009,Gaiotto:2009c}, which in turn also agrees with the  number of parameters in the proposed five-brane web description of $T_N$ \cite{Benini:2009}. 

\setcounter{equation}{0}
\section{B-model approach: The topological recursion}  \label{sB}

In this section we show how the 
topological recursion method, applied to Gaiotto's version of the Seiberg-Witten curve equipped with 
the M-theory differential, generates the semiclassical expansion of conformal blocks
 in the Liouville theory.

We already mentioned in section \ref{srev} that our motivation to employ the B-model approach 
comes from  the topological string engineering setup. 
However, applying the topological recursion in this context is fairly natural even without appealing to the relation to topological strings and  to the five-dimensional  
lift.
Since the insertion of a degenerate operator in the Liouville theory
is captured, in the semiclassical limit,  by  the integral of the Seiberg-Witten differential (\ref{Gz}), it is  natural to expect that multiple insertions should correspond to integrals of multiple differentials on the Seiberg-Witten curve $\Sigma$.

The topological recursion  is the natural formalism to generate multiple differentials on algebraic curves \cite{Eynard:2007,Eynard:2008we}.
This method was developed in the context of matrix models as a new way of solving loop equations using the spectral curve of the matrix model; however, the recursion turns out to be quite general and it can be use to generate multiple differentials (and closed symplectic invariants)
on arbitrary algebraic curves.

The basic ingredients of the recursion method are an algebraic curve equipped with a differential together with the Bergmann kernel $B(p,q)$ ---  the unique double differential on the curve with a double pole at $p=q$ (with residue one).

The topological recursion generates a set of multiple meromorphic differentials
$\mathcal{W}^{(g)}_k(p_1\cdots p_k)\rd p_1 \cdots \rd p_k $ of genus $g$.
We conjecture that the semiclassical expansion in 
powers of $\hbar$  of the conformal blocks with $k$ degenerate operator insertions can be written in terms of  the genus $g$ open ($k$-point)
B-model amplitudes:
\be
\label{genearl}
A^{(g)}_k(z_1,\cdots,z_k)=\int^{z_1}\!\! \cdots\int^{z_k} 
\rd p_1 \cdots \rd p_k \mathcal{W}^{(g)}_k(p_1,\cdots,p_k)\,.
\ee
For the case of one insertion we have the following expression\footnote{Note that this is the expansion of a determinant, see e.g.~(3.35), (3.36) in \cite{Marino:2007}. This is perfectly consistent with the fact that, as mentioned in section \ref{srev}, the insertion of a degenerate operator in the $2d$ CFT corresponds to the insertion of a determinant in the matrix model language of \cite{Dijkgraaf:2009}.} 
\bea \label{tr}
Z_{\rm null}(z)\big|_{\mathcal{Q}=0}&=&\exp\Big[\sum_{g,k} \hbar^{2g-2+k} \frac{1}{k!}A^{(g)}_k(z,\cdots,z) \Big] \\ 
&=&
\exp\Big[\frac{1}{\hbar} A^{(0)}_1(z){+} \frac{1}{2!}A^{(0)}_2(z,z){+}\hbar\left(A^{(1)}_1(z)+\frac{1}{3!} A^{(0)}_3(z,z,z)\right)+\cdots\Big].\nonumber
\eea
By comparing to the CFT expressions, this  in particular implies that the following equalities 
 should hold
\be
\label{3checks}
G_{0}(z)\big|_{\mathcal{Q}=0}=A^{(0)}_1(z),\;\; G_{1}(z)\big|_{\mathcal{Q}=0}= \frac{1}{2!}A^{(0)}_2(z,z),\;\;
G_{2}(z)\big|_{\mathcal{Q}=0}=A^{(1)}_1(z)+\frac{1}{3!} A^{(0)}_3(z,z,z)\,.
\ee
We will provide non-trivial checks of these relations below for the $\mathcal{T}_{3,0}(A_1)$ and  $\mathcal{T}_{4,0}(A_1)$ theories. 

For multiple insertions the only change in the above expression (\ref{tr}) is
\be
A^{(g)}_j(z,\cdots,z)  \rar \sum_{i_1,\ldots,i_j=1}^k A^{(g)}_j(z_{i_1},\cdots,z_{i_j})\,.
\ee
This more general expression will also be compared to the CFT expressions below. For instance, we will verify that 
\bea
\label{check2}
G_{1}(z_1,z_2)\big|_{\mathcal{Q}=0}=A^{(0)}_2(z_1,z_2) + \half(A^{(0)}_2(z_1,z_1) +A^{(0)}_2(z_2,z_2)  ) \,.
\eea

We note that the form of the expansion (\ref{tr}) is closely related to a `Baker-Akhiezer' function \cite{Eynard:2007,bbt} of the classical integrable system\footnote{We thank N. Orantin for a useful discussion on this point.} $\Psi(z)=(\psi_1\cdots\psi_d$)
\bea
\label{BA}
\psi(z)_i&=&\exp\Big[\sum_{g,k} \hbar^{2g-2+k} \frac{1}{k!}\int_{\nu_i}^z\cdots \int_{\nu_i}^z  
\rd p_1 \cdots \rd p_k \mathcal{W}^{(g)}_k(p_1,\cdots,p_k) \Big]\\&=&
\exp\Big[\frac{1}{\hbar} A^{(0)}_1(z) + \frac{1}{2!} A^{(0)}_2(z,z)+\hbar\left(A^{(1)}_1(z)+\frac{1}{3!} A^{(0)}_3(z,z,z)\right)+\cdots\Big], \nonumber
\eea
where $\nu_i$ are punctures of the differential.

In this case the Seiberg-Witten curve has genus zero and for this case it has been shown in \cite{Eynard:2007} that the  Baker-Akhiezer function
satisfies a Hirota equation (for the multi-component KP hierarchy)\footnote{See also \cite{Eynard:2008he} for the higher  genus case.}. It is natural to speculate that $Z_{\rm null}(z)$ for generic $\cQ$ will provide the Baker-Akhiezer function for a certain
$\cQ$uantum integrable system.
The $\beta$--ensemble formalism may allow one to compute the $G^k_i(z_1\cdots z_k)$ terms in  the $\cQ$ expansion (\ref{gqexp}) and  possibly  provide a way to `quantize' the integrable system with a B-model approach
\cite{Chekhov:2009mm,Eynard:2008mz}. 

It would be  interesting to explore the relation of the $\beta$-ensemble quantization to the recent work \cite{Nekrasov:2009b} (see also \cite{Nekrasov:2010}),
where it was  proposed and tested in some examples, that given a classical integrable system associated with a $\mathcal{N}=2$ gauge theory, its quantization  is provided by the Nekrasov equivariant partition function.

\subsection{Surface operators in the $T_2$ theory} 

The  Seiberg--Witten curve of the  $T_2$ theory is 
a  genus zero Riemann surface and the Seiberg-Witten one form reads, cf.~(\ref{T2SW}):
\be
\lambda_{SW}= x \, \D z = \frac{\sqrt{\alpha_1^2(1-z)+\alpha_0^2 (z^2-z)+\alpha_2^2 z}}{z(z-1)}\, \rd z\equiv M(z)\sqrt{\sigma(z)}\,\rd z,
\ee
where we have introduced the notation
\be
M(z)=\frac{\alpha_0}{z(z-1)}\,,\qquad\sqrt{\sigma(z)}=(z-\lambda_1)(z-\lambda_2)\,.
\ee

The Bergmann kernel for a genus zero hyper-elliptic Riemann surface
\cite{Ak} is given by:
\be
B(z_1,z_2)=\frac{1}{2(z_1-z_2)^2}\Big[\frac{z_1 z_2 -\frac{1}{2}(z_1+z_2)(\lambda_1+\lambda_2)+\lambda_1\lambda_2 }{\sqrt{(z_1-\lambda_1)(z_2-\lambda_1)(z_2-\lambda_1)(z_2-\lambda_2)}}+1\Big]\rd z_1 \rd z_2\,.
\ee
We will also need the kernel differentials for the genus zero case \cite{Ak,Bouchard:2007}:
\be
\chi_i^{(n)}(z_1)={1\over  (n-1)!} {1\over  {\sqrt {\sigma(z_1)}}} {\rd^{n-1} \over \rd z_2^{n-1}} \Biggl[{1\over 2 M(z_2) } {1\over z_1-z_2} \Biggr]_{z_2=\lambda_i}
\ee
Using the above ingredients, we can then use the topological recursion to generate multi--differentials. For example, the genus zero one-form is simply $\mathcal{W}^{(0)}_1(z_1)\, \D z_1=\la_{\mathrm SW}(z_1)$, the genus zero double differential is:
\be
\mathcal{W}^{(0)}_2(z_1,z_2)\, \rd z_1 \rd z_2 = B(z_1,z_2)-\frac{\rd z_1 \rd z_2}{(z_1-z_2)^2}\,,
\ee
the genus zero triple differential is:
\bea
\mathcal{W}^{(0)}_3(z_1,z_2, z_3)\, \rd z_1 \rd z_2\rd z_3&=& \half \sum_{i=1}^{2} M^2(\lambda_i) \sigma'(\lambda_i) \chi^{(1)}_i(z_1) \chi^{(1)}_i(z_2) \chi^{(1)}_i(z_3) \rd z_1 \rd z_2\rd z_3\,,\nonumber\\
\eea
and the genus one one-form is:
\bea
\mathcal{W}^{(1)}_1(z)\, \rd z &=&{1\over 16} \sum_{i=1}^{2} \chi^{(2)}_i(z)\,  \rd z -{1\over 8} \sum_{i=1}^{2} \biggl(\sum_{j\not=i} {1\over \lambda_i -\lambda_j} \biggr) \chi_i^{(1)}(z)\, \rd z\,.
\eea

The open amplitudes corresponding to the above differentials are as follows.

\noindent One-point function\footnote{This expression was also computed in \cite{Schiappa:2009} in the context of matrix models where it is known as the holomorphic effective potential.}:
\be
\label{a01}
A^{(0)}_1(z)=\alpha_1\log(z)+\frac{-\alpha_0^2+\alpha_1^2+\alpha_2^2}{2 \alpha_1}z-\frac{(\alpha_0^4 {-} 3 \alpha_1^4 {-} 6 \alpha_1^2 \alpha_2^2 {+} \alpha_2^4 {+} 2 \alpha_0^2 (\alpha_1^2 {-} \alpha_2^2))}{16 \alpha_1^3}z^2+\cdots
\ee
Two-point function:
\bea
\label{a02}
 A^{(0)}_2(z_1,z_2)&=&\frac{\alpha_0^4+(\alpha_1^2-\alpha_2^2)^2-2\alpha_0^2(\alpha_1^2+\alpha_2^2)}{16 \alpha_1^4} z_1 z_2+\\&+&
\nonumber
\frac{(\alpha_0^2+ \alpha_1^2-\alpha_2^2)
(\alpha_0^4+(\alpha_1^2 -\alpha_2^2)^2-2\alpha_0^2(\alpha_1^2+\alpha_2^2))
}{32 \alpha_1^6}
(z_1^2 z_2+z_1 z_2^2)+\cdots \non
\eea
Three-point function:
\bea
\label{a03}
A^{(0)}_3(z_1,z_2,z_3)&=&
\frac{(\alpha_0^2 - \alpha_2^2)(\alpha_0^4+(\alpha_1^2-\alpha_2^2)^2-2\alpha_0^2(\alpha_1^2+\alpha_2^2)}{ 8\alpha_1^{7}} z_1 z_2 z_3+\cdots
\eea
One-point function at genus one:
\bea
\label{a11}
 A^{(1)}_1(z)&=&\frac{\alpha_0^4+(\alpha_1^2-\alpha_2^2)^2-2\alpha_0^2(\alpha_1^2+\alpha_2^2)}{32 \alpha_1^5} z^2+\\&+&
\frac{(3\alpha_1^2+5(\alpha_0^2-\alpha_2^2)^2)(\alpha_0^4+(\alpha_1^2-\alpha_2^2)^2-2 \alpha_0^2(\alpha_1^2+\alpha_2^2)}{96 \alpha_1^7}z^3+\cdots \non
\eea
We have checked that the amplitudes (\ref{a01})-(\ref{a11}) match with the corresponding CFT results. For example, (\ref{a02}) is easily seen to be consistent with (\ref{g02}) using (\ref{check2}) and identifying $\al_3 \leftrightarrow \al_0$. Similarly, (\ref{a01}) agrees with (\ref{g01}).

The general pattern should be clear from the above examples. 
In conclusion, we find strong support in favour of the idea that the effects of multiple surface operator insertions can be calculated using topological recursion.

We conclude this section with a comment about the the quantum curve 
$\hat\psi_2(z)$ that we introduced in eq.~(\ref{qc}).
We can clearly apply the topological recursion using the quantum curve and compute
the open amplitudes $\hat A^{(g)}_h$ with expansion:
\bea
\hat{A}^{(g)}_k(z_1,\cdots z_k)=\sum_n \cQ^n A^{(g,n)}_k(z_1,\cdots z_k) \,.
\eea
 Is it easy to realize that they are obtained from the classical ones by simply replacing:
\be
\alpha_i^2 \to - (\cQ-\alpha_i)\alpha_i.
\ee
We checked that the linear terms in $\cQ$, $A^{(g,1)}_k(z_1,\cdots z_k)$ match with the corresponding terms in the  expansion of the $ G_i(z_1\cdots z_k)$. However,   
the higher order terms do not agree.
The $\cQ$ expansion of  $ G_i(z_1\cdots z_k)$ has a very
appealing `geometrical' feature since powers of $\cQ$ and the number of `boundaries' are correlated:  $k$--boundary amplitudes  have $\cQ^n$ contributions  with $n\leq k$.
This feature is not shared by the $\hat{A}^{(g)}_k(z_1,\cdots z_k)$.
We do not understand the origin of the mismatch. 
It is possible that one needs  to redefine  the open moduli with $\cQ$
in order to match the two expansions.

\subsection{Surface operators in the $\SU(2)$ theory with $N_f=4$} 

We will now use the topological recursion to compute amplitudes\footnote{An alternative (possibly related) expansion method based on the corresponding CFT expression was developed in \cite{Fateev:2009}. } for the  $\SU(2)$ theory with $N_f=4$ and compare them to the results in Liouville theory of section \ref{sCFT}.

The Seiberg-Witten differential is given by (see e.g.~\cite{Eguchi:2009}):
\be
\label{swp}
\lambda_{SW}= \frac{\sqrt{P_4(z)}\, \rd z}{ z (z-1)(z-\zeta)} \,,
\ee
with
\be
P_4(z)=m_0^2\left(z^4+z^3 S_1+z^2 S_2+z S_3+S_4\right)=\prod_{i=1}^4(z-\lambda_i),
\ee
where we introduced the symmetric combinations of the roots:
\be
S_k = \sum_{1 \leq j_1 < j_2 < \ldots < j_k \leq 4} \lambda_{j_1} \cdots \lambda_{j_k}\,.
\ee
More explicitly we have
\bea
&&S_1=-\frac{m_0^2+m_2^2(\zeta-1)+m_0^2 \zeta+2 m_2 m_3\zeta +(1+\zeta )U }{m_0^2}\,,\qquad  S_4=\frac{m_1^2 \zeta^2}{m_0^2}\,, \non \\
&&S_2=\frac{(m_0^2+m_1^2-m_3^2 +2 m_2 m_3)\zeta+m_2^2 (\zeta-1)\zeta+2 m_2 m_3\zeta^2 +m_3^2\zeta^2+(1+\zeta )^2U }{m_0^2},\nonumber\\
&&S_3=-\frac{(m_1^2-m_3^2)\zeta+(m_1^2+2m_2 m_3+m_3^2)\zeta^2+\zeta(1+\zeta )U }{m_0^2}\,.
\eea

One can check that the residues of the double poles of $\lambda_{SW}^2$ at $z=0,1,\zeta,\infty$
are given by the $m_i^2$.

To apply the topological recursion we need  the annulus amplitude.
In this case the Seiberg-Witten curve has genus one and the annulus amplitudes  can be written in terms of the Weierstrass elliptic function; alternatively
we can use an expression in terms of the branch points of the hyper--elliptic curve  due to Akemann \cite{Ak} which reads:
\bea
\mathcal{W}^{(0)}_2(z_1,z_2)\, \rd z_1 \rd z_2 &=& \frac{m_0^2 \,\rd z_1 \rd z_2}{4 \sqrt{P_4(z_1)P_4(z_2)}}\frac{M(z_1,z_2)+M(z_2,z_1)}{(z_1-z_2)^2} \\
&-& \frac{m_0^2 \, \rd z_1 \rd z_2}{4 \sqrt{P_4(z_1)P_4(z_2)}} (\lambda_1 - \lambda_3)(\lambda_2-\lambda_4) \frac{E(k)}{K(k)} - \frac{\rd z_1 \rd z_2}{2 (z_1-z_2)^2}\,,
\label{e:ak} \non
\eea
where
\be
M(z_1,z_2)=(z_1-\lambda_1)(z_1-\lambda_2)(z_2-\lambda_3)(z_2-\lambda_4)\,,
\ee
and
\be
k^2 = \frac{(\lambda_1-\lambda_2)(\lambda_3-\lambda_4)}{(\lambda_1-\lambda_3)(\lambda_2-\lambda_4)}\,.
\ee

Note that the amplitudes depend on a choice of ordering of the branch points, which corresponds to a choice of canonical basis of cycles on the Seiberg-Witten curve
that is to a  choice of  duality frame for the gauge theory.
A reordering of the branch points corresponds to a generalized 
$S$--duality transformation in the gauge theory.
Since we are interested in matching to the results of section 
\ref{sCFT} where the conformal blocks are perturbative in $\zeta$,
and correspond to a particular ordering of the vertex operators, we need  to 
expand the gauge theory in the duality frame 
corresponding to the small  $\zeta$ expansion. 

We also need the (inverse) mirror map to express the parameter 
$U$ in terms of $\zeta$ and $a$.
This can be obtained as follows.
By differentiating $\lambda_{SW}$ w.r.t.~the modulus  $U$ we obtain the holomorphic one--form:
\be
\omega\equiv\frac{\rd \lambda_{SW }}{\rd U}=-\frac{(1+\zeta) \rd z}{2 \sqrt{P_4(z)}}\,,
\ee
and its  $\alpha$-cycle integral:
\be
\label{dadU}
\frac{\rd a}{\rd U}=\frac{1}{2 \pi \ri}\oint_{\lambda_1,\lambda_2} \omega=
-\frac{1 }{  \pi \ri}  \frac{(1+\zeta)}{m_0\sqrt{(\lambda_2-\lambda_3)(\lambda_1-\lambda_4) }}
K(k^2)\,.
\ee
By inverting we obtain  the inverse  mirror map:
\be
U=a^2  +\zeta \left(-\frac{3}{2} a^2+\frac{(m_0^2{+}m_1^2{-}m_2^2{-}m_3^2{-}4 m_2 m_3)  }{2 }+\frac{(m_0^2{-}m_2^2)(m_1^2{-}m_3^2)}{2  a^2}\right)+\mathcal{O}(\zeta^2)\,.
\ee
We now have all the ingredients to compute the  one-point amplitude:
\bea
\label{a01su2}
A^{(0)}_1(z) &=&\frac{(a^2-m_0^2+m_2^2)z}{2 a}+\frac{(3 a^4 - 2 a^2 m_0^2 - m_0^4 + 6 a^2 m_2^2 + 2 m_0^2 m_2^2 - m_2^4)}{16 a^3}\,z^2 \nonumber\\&+&
\! \!  \zeta\Big[
\frac{(a {-} m_0 {-} m_2) (a {+} m_0 {-} m_2) (a {-} m_0 {+} m_2) (a {+} m_0 {+} m_2) (a^2 - m_1^2 +  m_3^2)}{16 a^5}\,z \nonumber\\
&+&\frac{m_1^2-m_3^2-a^2}{2az} +\cdots
\Big]+\mathcal{O}(\zeta^2)
\eea
and the two-point amplitude:
\bea
\label{a02su2}
\nonumber A^{(0)}_2(z_1,z_2) &=&
\frac{
\left(a^4-2a^2(m_0^2+  m_2^2)+(m_0^2-m_2^2)^2\right)
}{16 a^4}z_1z_2+\cdots\nonumber \\
&+&\zeta\Big[
\frac{ (a^2{-}m_1^2{+}m_3^2) (a^4{+}(m_0^2{-}m_2^2)^2{-}2a^2(m_0^2{+}m_2^2)) (m_2^2{-}m_0^2)}{16 a^8} z_1 z_2+ \cdots\Big] \nonumber 
\\ &+&\mathcal{O}(\zeta^2).
\eea
By taking into account the dictionary (when $\cQ=0$):
\be
m_0=\al_4  \,,\quad m_1= \al_1\,, \quad m_2 = \al_3\,,\quad m_3= \al_2\,, \quad a=\si\,,
\ee
we have matched these expressions to the corresponding CFT results; for instance, (\ref{a02su2}) is easily seen to agree with (\ref{g02su2}).

\setcounter{equation}{0}
\section{A-model approach: Toric branes and the topological vertex} \label{sA}

\par{Type II string theory (or M-theory) can be compactified on a Calabi-Yau threefold to engineer  a  supersymmetric gauge theory on the transverse $4d$ (or $5d$) space. Of special interest are the toric geometries since they engineer ${\cal N}=2$ supersymmetric (quiver) gauge theories with $\SU(N)$ gauge group(s) with or without matter fields in different representations. Topological string theory on toric Calabi-Yau threefolds provide a geometric way of computing the prepotential and higher-genus gravitational corrections in these theories. 
The topological vertex formalism allows one to perform 
all genus computations of topological string amplitudes on toric geometries. The trivalent topological vertex \cite{Aganagic:2003a} is defined in terms of the open topological string amplitude on ${\mathbb C}^{3}$. The partition function for a generic toric Calabi-Yau threefold can be computed by decomposing the geometry into ${\mathbb C}^{3}$ patches. Then an appropriate gluing algorithm allows one to obtain the full result using the individual contributions coming from each patch. Pictorially, each patch corresponds to a trivalent vertex in the so-called toric diagram, which encodes the degenerating 2-cycles of the geometry. 
For further details of this construction and the gluing algorithm we refer  to the original work \cite{Aganagic:2003a}. Here we just give the explicit form of the topological vertex, which is labeled by three irreducible representations $\lambda$, $\mu$ and $\nu$ of the $\U(\infty)$ algebra, each  corresponding to one of the three legs of the vertex: }
\bea \label{topvertex}
C_{\lambda\mu\nu}(q)=q^{\frac{\kappa(\mu)}{2}}s_{\nu^{t}}(q^{-\rho})\sum_{\eta}s_{\lambda^{t}/\eta}(q^{-\nu-\rho})s_{\mu/\eta}(q^{-\nu^{t}-\rho}) \,.
\eea
In this expression $s_{\lambda}$ and $s_{\lambda/\eta}$ denote the Schur and skew-Schur function, respectively. The arguments of the Schur functions involve $q^{-\nu-\rho}=\{q^{-\nu_{i}+i-1/2}\}_{i=1}^{\infty}$. 

There exists an extension of topological vertex formalism \cite{Iqbal:2007,Awata:2005} that  computes the {\em refined} topological string partition functions on toric geometries. This construction was motivated by Nekrasov's computation of the instanton partition function and by his conjecture of the relationship of this quantity to topological string partition functions \cite{Nekrasov:2002}. According to this conjecture, when the equivariant parameters $\epsilon_{1,2}$  in the instanton partition function sum up to zero, the remaining variable can be identified with the topological string coupling constant $g_{s}$, i.e.~$\epsilon_{1}=-\epsilon_{2}=g_{s}$, and with this identification the  topological string partition function agrees with the instanton partition function (in $d=5$). 

The derivation of the refined topological vertex is based on the combinatorial interpretation of the topological vertex and has the following explicit form \cite{Iqbal:2007}\footnote{An alternative form was proposed in \cite{Awata:2005}.  }:

\bea \nonumber
C_{\lambda\mu\nu}(t,q)=\left(\frac{q}{t} \right)^{\frac{\Arrowvert\mu\Arrowvert^{2}+\Arrowvert\nu\Arrowvert^{2}}{2}} \!\!  t^{\frac{\kappa(\mu)}{2}}P_{\nu^{t}}(t^{-\rho};q,t)\sum_{\eta}\left(\frac{q}{t}\right)^{\frac{|\eta|+|\lambda|-|\mu|}{2}}  \!\!\!\!  s_{\lambda^{t}/\eta}(t^{-\rho}q^{-\nu})s_{\mu/\eta}(t^{-\nu^{t}}q^{-\rho}). \\
\eea 
where $q=e^{\epsilon_{1}}$, $t=e^{-\epsilon_{2}}$ and $P_{\nu}(\mathbf{x};q,t)$ is the Macdonald function.  It is known that $P_{\nu}(q^{-\rho};q,q)=s_{\nu}(q^{-\rho})$. Using this fact it is easy to see that the refined topological vertex reduces to the usual vertex (\ref{topvertex}) in the special case $t=q$. Unlike the usual vertex, the refined vertex does not possess a cyclic symmetry in the three representations labelling its legs. Instead it has a so-called preferred direction which is chosen to be along the last leg, labelled by $\nu$. The gluing rules are the same as the ones for the usual vertex as long as one pays attention to the preferred direction (see \cite{Iqbal:2007} for further details). The refined  topological string partition function agrees with the $5d$ instanton partition function with unconstrained $\ep_{1,2}$ \cite{Iqbal:2007,Awata:2005,Taki:2007}.

For toric geometries whose toric diagrams are triangulations of a `strip', a very useful computational  tool has been  developed in \cite{Iqbal:2004}. We summarise this method in appendix \ref{striprules}.

\subsection{Toda theory (chiral) three-point functions from toric geometry?}

In \cite{Benini:2009} it was proposed that a five-dimensional version of the $T_N$ ($\cT_{3,0}(A_r)$) theory  can be obtained in a IIB setup  
in terms of a certain $N$-junction. This $N$-junction is a particular configuration (web-diagram) of $N$ D5-branes, $N$ NS5-branes and $N$ $(1,1)$ 5-branes. For each web-diagram one can also form a dual diagram by exchanging vertices with faces. These diagrams turn out to be the toric diagrams of a particular toric CY threefold (a blow up of the $\mathbb{C}^3/[\mathbb{Z}_N{\times} \mathbb{Z}_N]$ orbifold). 

We will argue that the toric interpretation of the web diagram should be taken seriously, implying that topological string techniques can be used to analyse these theories. Below we will collect some evidence in favour of this idea.  As an example, consider the $N=2$ case, i.e.~the $T_2$ theory. In this case the toric diagram takes the following form:

\begin{figure}[h]
\begin{center}
\psset{unit=0.8cm}
\begin{pspicture}(7,7)
\psline[linecolor=red,linestyle=solid,linewidth=1pt]{-}(1,1)(7,1)
\psline[linecolor=red,linestyle=solid,linewidth=1pt]{-}(1,1)(1,7)
\psline[linecolor=red,linestyle=solid,linewidth=1pt]{-}(1,7)(7,1)
\psline[linecolor=red,linestyle=solid,linewidth=1pt]{-}(1,4)(4,4)
\psline[linecolor=red,linestyle=solid,linewidth=1pt]{-}(4,4)(4,1)
\psline[linecolor=red,linestyle=solid,linewidth=1pt]{-}(1,4)(4,1)

\psline[linecolor=black,linestyle=solid,linewidth=2pt]{-}(2,2)(3,3)
\psline[linecolor=black,linestyle=solid,linewidth=2pt]{-}(2,2)(2,0)
\psline[linecolor=black,linestyle=solid,linewidth=2pt]{-}(2,2)(0,2)
\psline[linecolor=black,linestyle=solid,linewidth=2pt]{-}(3,3)(4.5,3)
\psline[linecolor=black,linestyle=solid,linewidth=2pt]{-}(4.5,3)(4.5,0)
\psline[linecolor=black,linestyle=solid,linewidth=2pt]{-}(4.5,3)(5.5,4)
\psline[linecolor=black,linestyle=solid,linewidth=2pt]{-}(3,3)(3,4.5)
\psline[linecolor=black,linestyle=solid,linewidth=2pt]{-}(3,4.5)(4,5.5)
\psline[linecolor=black,linestyle=solid,linewidth=2pt]{-}(3,4.5)(0,4.5)

\end{pspicture}
\caption{The toric diagram for the  $T_2$ geometry.} \label{webs1}
\end{center}
\end{figure}

As mentioned in section \ref{Toda3}, the  $T_N$ theories should presumably be related, via the AGT conjecture, to (chiral) three-point functions in the $A_{N-1}$ Toda theories. Therefore, the toric point of view may offer a possible way to compute (chiral) Toda theory three-point functions
as partition functions of the topological string on the
 toric $\mathbb{C}^3/\mathbb{Z}_N{\times} \mathbb{Z}_N$ geometries. 

As a first check we analyse the $T_2$ case using  the refined topological vertex to compute the topological string partition function of the toric geometry in figure \ref{webs1}. The partition function for this geometry in the unrefined case was first computed in \cite{Karp:2005} (see also~\cite{Sulkowski:2006}).

\begin{figure}[h]
\begin{center}
\psset{unit=1.5cm}
\begin{pspicture}(4,4)

\psline[linecolor=black,linestyle=solid,linewidth=2pt]{-}(0,1)(2,1)
\psline[linecolor=black,linestyle=solid,linewidth=2pt]{-}(2,1)(3,0)
\psline[linecolor=black,linestyle=solid,linewidth=2pt]{-}(2,1)(2,2)
\psline[linecolor=black,linestyle=solid,linewidth=2pt]{-}(2,2)(1,3)
\psline[linecolor=black,linestyle=solid,linewidth=2pt]{-}(1,3)(1,4)
\psline[linecolor=black,linestyle=solid,linewidth=2pt]{-}(1,3)(0,3)
\psline[linecolor=black,linestyle=solid,linewidth=2pt]{-}(2,2)(3,2)
\psline[linecolor=black,linestyle=solid,linewidth=2pt]{-}(3,2)(4,1)
\psline[linecolor=black,linestyle=solid,linewidth=2pt]{-}(3,2)(3,4)

\psline[linecolor=blue,linestyle=solid,linewidth=2pt]{-}(1.5,2.5)(1.7,2.7)
\psline[linecolor=blue,linestyle=solid,linewidth=2pt]{-}(1.5,2.5)(1.3,2.3)
\psline[linecolor=blue,linestyle=solid,linewidth=2pt]{-}(2.5,0.5)(2.3,0.3)
\psline[linecolor=blue,linestyle=solid,linewidth=2pt]{-}(2.5,0.5)(2.7,0.7)
\psline[linecolor=blue,linestyle=solid,linewidth=2pt]{-}(3.5,1.5)(3.3,1.3)
\psline[linecolor=blue,linestyle=solid,linewidth=2pt]{-}(3.5,1.5)(3.7,1.7)

\put(1.7,2.4){$\lambda,Q_{1}$}
\put(1.3,1.5){$\nu,Q_{3}$}
\put(2.3,1.7){$\mu,Q_{2}$}

\end{pspicture}
\caption{Closed topological vertex with a choice of preferred direction.} \label{vertexxxx}
\end{center}
\end{figure}

Using the notation and preferred direction as in figure \ref{vertexxxx}, the partition function for the $T_2$ geometry, sometimes called the closed topological vertex, can be computed\footnote{The  equivalence of the product representation with the refined vertex computation has been verified up to  fifth order in each K\"{a}hler parameter with the aid of a computer code. Our expression also reduces to the result in \cite{Karp:2005} when $q=t$ as required for consistency.}

\bea\nonumber
\label{t2}
Z'_{\rm T_2}\!\!\!\!\!&=&\!\!\!
\!\sum_{\lambda}(-Q_{1})^{|\lambda|}q^{\frac{\Arrowvert\lambda\Arrowvert^{2}}{2}} t^{\frac{\Arrowvert\lambda^{t}\Arrowvert^{2}}{2}}\widetilde{Z}_{\lambda}(t,q)\widetilde{Z}_{\lambda^{t}}(q,t)\prod_{i,j=1}^{\infty}\frac{(1{-}Q_{2}\,q^{-\rho_{j}}t^{-\lambda_{j}^{t}-\rho_{i}})(1{-}Q_{3}\,q^{-\lambda_{j}-\rho_{i}}t^{-\rho_{j}})}{(1-Q_{2}Q_{3}\,q^{-\rho_{i}-1/2}t^{-\rho_{j}+1/2})} \\
\nonumber
\!\!\!\!\!&=&\!\!\!
\!
\prod_{i,j=1}^{\infty}\frac{(1-Q_{1}\,q^{-\rho_{i}}t^{-\rho_{j}})(1-Q_{2}\,q^{-\rho_{i}}t^{-\rho_{j}})(1-Q_{3}\,q^{-\rho_{i}}t^{-\rho_{j}})(1-Q_{1}Q_{2}Q_{3}\,q^{-\rho_{i}}t^{-\rho_{j}})}{(1{-}Q_{1}Q_{2}\,q^{-\rho_{i}+\frac{1}{2}}t^{-\rho_{j}{-}\frac{1}{2}})(1{-}Q_{1}Q_{3}\,q^{-\rho_{i}{+}\frac{1}{2}}t^{-\rho_{j}-\frac{1}{2}})(1{-}Q_{2}Q_{3}\,q^{-\rho_{i}-\frac{1}{2}}t^{-\rho_{j}+\frac{1}{2} })},
\\\!\!\!\!\!&=&\!\!\!
\!
\prod_{i,j=1}^{\infty}\frac{(1{-}Q_{1}Q_{2}\,q^{\rho_{i}+\frac{1}{2}}t^{-\rho_{j}{-}\frac{1}{2}})(1{-}Q_{1}Q_{3}\,q^{\rho_{i}{+}\frac{1}{2}}t^{-\rho_{j}-\frac{1}{2}})(1{-}Q_{2}Q_{3}\,q^{\rho_{i}-\frac{1}{2}}t^{-\rho_{j}+\frac{1}{2} })}{(1-Q_{1}\,q^{\rho_{i}}t^{-\rho_{j}})(1-Q_{2}\,q^{\rho_{i}}t^{-\rho_{j}})(1-Q_{3}\,q^{\rho_{i}}t^{-\rho_{j}})(1-Q_{1}Q_{2}Q_{3}\,q^{\rho_{i}}t^{-\rho_{j}})},
\eea
where in the final equality we have used the analytic continuation
\be
 \prod_{i,j=1}^{\infty}(1- Q q^{-\rho_{i}}
t^{-\rho_{j}})=\prod_{i,j=1}^{\infty}(1-Q q^{-\rho_{i}}t^{\rho_{j}})^{-1}.
\ee
The  product expressions in (\ref{t2}) have a clear interpretation in terms of contributions coming from wrapping various combinations of the three 2-cycles with K\"ahler classes $Q_i$.

There exists an alternative way to obtain topological string partition functions using a relation to statistical models of crystal melting \cite{Okounkov:2003}. In this approach one also obtains extra multiplicative factors involving the MacMahon function related to the constant map contribution, not included in the vertex. In the crystal approach to the closed topological vertex discussed in \cite{Sulkowski:2006} it was found that the topological vertex computation should be multiplied by  a single MacMahon factor in order to agree with the crystal computation. In our computation one should therefore supplement the above result (\ref{t2}) with a factor corresponding to a refined version of the MacMahon function. There are different proposals for this function in the literature, all of the form
\be \label{MacM}
M(q,t) = \prod_{i,j=1}^{\infty} (1- \,q^{\rho_{i}+\frac{\de}{2} }t^{\rho_{j}-\frac{\de}{2} })^{-1}
=  \prod_{i,j=1}^{\infty} (1 - \,q^{\rho_{i}+\frac{\de}{2} }t^{-\rho_{j}-\frac{\de}{2} })
\ee
In \cite{Iqbal:2007} the choice $\de=-1$ was made, whereas in the later work \cite{Dimofte:2009}  $\de=0$ was argued to be natural. 

We now want to relate $Z_{\rm T_2}= M(q,t)  Z'_{\rm T_2}$ to a (chiral) three-point function in 
 the Liouville theory. The first thing to note is that  $Z_{\rm T_2}$  corresponds in gauge theory language to a $5d$ theory and therefore should correspond to a $q$-deformed version 
 of the Liouville theory (see e.g.~\cite{Shiraishi:1995}). To obtain an expression that can be compared to the usual Liouville result  we need to take the four-dimensional limit.  A quick heuristic way to perform the reduction is the following. 
 
 We use   $Q=e^{-2 R m} $, $q^{\rho_i}=e^{ 2 R (i-\frac{1}{2}) \ep_1}$ and $t^{\rho_j}=e^{-2 R  (j-\frac{1}{2}) \ep_2}$ and write
\be \label{sinhrule}
\prod_{i,j=1}^{\infty}(1-Q \,q^{-\rho_{i}}t^{\rho_{j}}) = \prod_{i,j=1}^{\infty} e^{-R [m+  i\ep_1 + j \ep_2-\ep/2] } 2  \sinh(R[m + i \ep_1 + j \ep_2]).
\ee
Applying (\ref{sinhrule}) to all the factors in (\ref{t2}), (\ref{MacM}) taking the limit $R\rar 0$, and using 
\be
\Ga_2(x|\ep_1,\ep_2) \propto \prod_{i,j=0}^{\infty} (x + i \ep_1 + j \ep_2 )^{-1}  ,
\ee
where $\Ga_2(x|\ep_1,\ep_2)$ is the Barnes double gamma function, we find ($ \Ga_2(x)\equiv \Ga_2(x|\ep_1,\ep_2)$)
\be
\frac{ \Ga_2(m_1+\ep/2)\Ga_2(m_2+\ep/2)\Ga_2(m_3+\ep/2)\Ga_2(m_1+m_2+m_3+\ep/2) }{\Ga_2(\ep/2) \Ga_2(m_1+m_2+\ep)\Ga_2(m_1+m_3+\ep)\Ga_2(m_1+m_3+\ep)},
\ee
where we chose $\de=0$ in (\ref{MacM}). Using the following dictionary (as usual, $\ep=\cQ$):
\be
\label{dict}
m_1=-\al_1{+}\al_2{+}\alpha_3{-}\cQ/2, \quad m_2=\al_1{-}\al_2{+}\al_3{-}\cQ/2,\quad 
m_3=\al_1{+}\al_2{-}\al_3{-}\cQ/2\,,
\ee
finally leads to 
\bea \label{L3pt}
\frac{ \Ga_2(-\al_1{+}\al_2{+}\alpha_3)\Ga_2(\al_1{-}\al_2{+}\al_3)\Ga_2(\al_1{+}\al_2{-}\al_3)\Ga_2(\al_1{+}\al_2{+}\al_3{-}\cQ) }{\Ga_2(\cQ/2)\Ga_2(2\al_1)\Ga_2(2\al_2)\Ga_2(2\al_3)}\,.
\eea
This expression can be related to a chiral version of the Liouville three-point function. 
It can be shown that, after suitably rescaling the vertex operators  with multiplicative factors depending on their momenta, the Liouville three--point function \cite{Dorn:1994} can be written 
\be \label{cL}
  \big|\Ga_2( \al_1{+}\al_2{+}\al_3-\cQ) \Ga_2(-\al_1{+}\al_2{+}\al_3)\Ga_2(\al_1{-}\al_2{+}\al_3) \Ga_2(\al_1{+}\al_2{-}\al_3) \big|^2\,.
\ee
Hence there is a natural definition of a chiral three--point function in the Liouville theory as the ``square root" of (\ref{cL}). We see that this expression agrees with the numerator in (\ref{L3pt}).  
Let us also mention that in \cite{Schiappa:2009} the three-point function was analysed from the matrix model perspective, essentially obtaining the same expression as in (\ref{L3pt}) (modulo some subtle points that still need to be clarified).

Finally, we observe that if one chooses $\de=1$ in (\ref{MacM}), then the resulting expression for the partition function precisely agrees with the chiral three-point function in (6.21)-(6.22) in \cite{Teschner:2008}. 

\begin{figure}[h]
\begin{center}
\psset{unit=0.6cm}
\begin{pspicture}(10,10)

\psline[linecolor=red,linestyle=solid,linewidth=1pt]{-}(0,1)(0,10)
\psline[linecolor=red,linestyle=solid,linewidth=1pt]{-}(0,1)(9,1)
\psline[linecolor=red,linestyle=solid,linewidth=1pt]{-}(0,10)(9,1)
\psline[linecolor=red,linestyle=solid,linewidth=1pt]{-}(0,4)(6,4)
\psline[linecolor=red,linestyle=solid,linewidth=1pt]{-}(0,7)(3,7)
\psline[linecolor=red,linestyle=solid,linewidth=1pt]{-}(3,7)(3,1)
\psline[linecolor=red,linestyle=solid,linewidth=1pt]{-}(6,4)(6,1)
\psline[linecolor=red,linestyle=solid,linewidth=1pt]{-}(0,7)(6,1)
\psline[linecolor=red,linestyle=solid,linewidth=1pt]{-}(0,4)(3,1)
\psline[linecolor=black,linestyle=solid,linewidth=2pt]{-}(1.25,2.25)
(1.75,2.75)
\psline[linecolor=black,linestyle=solid,linewidth=2pt]{-}(1.75,2.75)
(1.75,4.9)
\psline[linecolor=black,linestyle=solid,linewidth=2pt]{-}(1.75,4.9)
(2.25,5.4)
\psline[linecolor=black,linestyle=solid,linewidth=2pt]{-}(1.25,2.25)
(1.25,0)
\psline[linecolor=black,linestyle=solid,linewidth=2pt]{-}(1.25,2.25)
(-1,2.25)
\psline[linecolor=black,linestyle=solid,linewidth=2pt]{-}(1.75,2.75)
(3.9,2.75)
\psline[linecolor=black,linestyle=solid,linewidth=2pt]{-}(3.9,2.75)
(4.4,3.25)
\psline[linecolor=black,linestyle=solid,linewidth=2pt]{-}(2.25,5.4)
(4.4,5.4)
\psline[linecolor=black,linestyle=solid,linewidth=2pt]{-}(4.4,5.4)
(4.4,3.25)
\psline[linecolor=black,linestyle=solid,linewidth=2pt]{-}(4.4,5.4)
(5.5,6.5)
\psline[linecolor=black,linestyle=solid,linewidth=2pt]{-}(3.9,2.75)
(3.9,0)
\psline[linecolor=black,linestyle=solid,linewidth=2pt]{-}(1.75,4.9)
(-1,4.9)
\psline[linecolor=black,linestyle=solid,linewidth=2pt]{-}(2.25,5.4)
(2.25,7.4)
\psline[linecolor=black,linestyle=solid,linewidth=2pt]{-}(2.25,7.4)
(-1,7.4)
\psline[linecolor=black,linestyle=solid,linewidth=2pt]{-}(2.25,7.4)
(3.45,8.60)
\psline[linecolor=black,linestyle=solid,linewidth=2pt]{-}(4.4,3.25)
(6.47,3.25)
\psline[linecolor=black,linestyle=solid,linewidth=2pt]{-}(6.47,3.25)
(7.6,4.38)
\psline[linecolor=black,linestyle=solid,linewidth=2pt]{-}(6.47,3.25)
(6.47,0)

\end{pspicture}
\caption{The toric diagram for the $T_3$ geometry.} \label{webs2}
\end{center}

\end{figure}
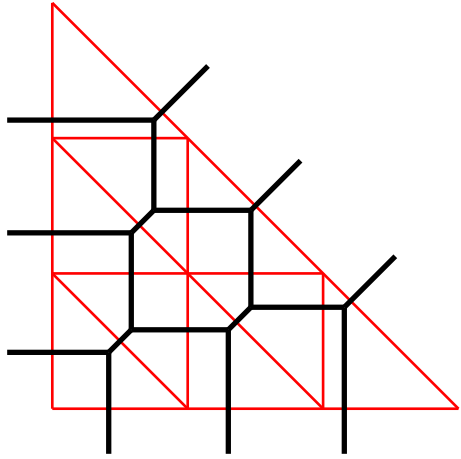

For the higher rank $T_N$ cases the situation is much more involved. For instance, consider the  $T_3$ theory. In this case the toric diagram is given in figure \ref{webs2}.

In the four-dimensional limit this theory should reduce \cite{Gaiotto:2009} to the $E_6$ theory first constructed in \cite{Minahan:1996}. Not much is known about this strongly-coupled theory (see however \cite{Gadde:2010} for a recent result). It is easy to see that the toric diagram in figure \ref{webs2} corresponds to an $\SU(2)$ theory with $N_f=5$  in $d=5$.  This connection is not surprising since it was argued in \cite{Seiberg:1996} that the $5d$ $\SU(2)$ theory with $N_f=5$ is related to a five-dimensional version of the $E_6$ theory. 

It is technically difficult to study the above toric $T_3$ geometry.  Another obstacle is that  the corresponding three-point functions on the Toda side are not known. But as a first consistency check one can verify that  the total  number of independent  K\"ahler classes in the toric geometry agrees with the counting of parameters in the Toda three-point functions in section \ref{Toda3}. 
Even though an explicit product formula is not possible (because of the presence of a non-trivial four-cycle in the geometry), one can still write down the vertex computation for the above geometry. But, it is not clear what (if anything) the result should correspond to in the $A_2$ Toda theory.  It may be that the patch in moduli space where the vertex computation is applicable is not the relevant patch for the comparison with the Toda result. 
Furthermore, to get results in the usual Toda theory (rather than in some $q$-deformed version) one should take the limit to four dimensions  which may be subtle for this theory.

\subsection{Some facts about the geometric engineering of $\SU(N)$ theories}

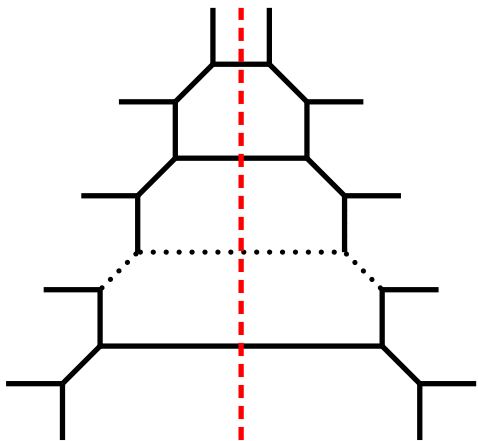
\begin{figure}[h]
\psset{unit=0.5cm}
\begin{center}
\begin{pspicture}(12.5,11.5)

\psline[linecolor=black,linestyle=solid,linewidth=2pt]{-}(0,1.5)
(1.5,1.5)
\psline[linecolor=black,linestyle=solid,linewidth=2pt]{-}(1.5,1.5)
(1.5,0)
\psline[linecolor=black,linestyle=solid,linewidth=2pt]{-}(1.5,1.5)
(2.5,2.5)
\psline[linecolor=black,linestyle=solid,linewidth=2pt]{-}(2.5,2.5)
(2.5,4)
\psline[linecolor=black,linestyle=dotted,linewidth=2pt]{-}(2.5,4)(3.5,5)
\psline[linecolor=black,linestyle=solid,linewidth=2pt]{-}(3.5,5)
(3.5,6.5)
\psline[linecolor=black,linestyle=solid,linewidth=2pt]{-}(3.5,6.5)
(4.5,7.5)
\psline[linecolor=black,linestyle=solid,linewidth=2pt]{-}(4.5,7.5)
(4.5,9)
\psline[linecolor=black,linestyle=solid,linewidth=2pt]{-}(4.5,9)(5.5,10)
\psline[linecolor=black,linestyle=solid,linewidth=2pt]{-}(5.5,10)
(5.5,11.5)
\psline[linecolor=black,linestyle=solid,linewidth=2pt]{-}(5.5,10)(7,10)
\psline[linecolor=black,linestyle=solid,linewidth=2pt]{-}(7,10)(7,11.5)
\psline[linecolor=black,linestyle=solid,linewidth=2pt]{-}(7,10)(8,9)
\psline[linecolor=black,linestyle=solid,linewidth=2pt]{-}(8,9)(8,7.5)
\psline[linecolor=black,linestyle=solid,linewidth=2pt]{-}(8,7.5)(9,6.5)
\psline[linecolor=black,linestyle=solid,linewidth=2pt]{-}(9,6.5)(9,5)
\psline[linecolor=black,linestyle=dotted,linewidth=2pt]{-}(9,5)(10,4)
\psline[linecolor=black,linestyle=solid,linewidth=2pt]{-}(10,4)(10,2.5)
\psline[linecolor=black,linestyle=solid,linewidth=2pt]{-}(10,2.5)
(11,1.5)
\psline[linecolor=black,linestyle=solid,linewidth=2pt]{-}(11,1.5)(11,0)
\psline[linecolor=black,linestyle=solid,linewidth=2pt]{-}(11,1.5)
(12.5,1.5)
\psline[linecolor=black,linestyle=solid,linewidth=2pt]{-}(2.5,2.5)
(10,2.5)
\psline[linecolor=black,linestyle=dotted,linewidth=2pt]{-}(3.5,5)(9,5)
(9,6.5)
\psline[linecolor=black,linestyle=solid,linewidth=2pt]{-}(4.5,7.5)
(8,7.5)
\psline[linecolor=black,linestyle=solid,linewidth=2pt]{-}(2.5,4)(1,4)
\psline[linecolor=black,linestyle=solid,linewidth=2pt]{-}(10,4)(11.5,4)
\psline[linecolor=black,linestyle=solid,linewidth=2pt]{-}(2,6.5)
(3.5,6.5)
\psline[linecolor=black,linestyle=solid,linewidth=2pt]{-}(10.5,6.5)
(9,6.5)
\psline[linecolor=black,linestyle=solid,linewidth=2pt]{-}(4.5,9)(3,9)
\psline[linecolor=black,linestyle=solid,linewidth=2pt]{-}(8,9)(9.5,9)

\psline[linecolor=red,linestyle=dashed,linewidth=2pt]{-}(6.25,0)
(6.25,11.5)

\end{pspicture}
\caption{The toric version of $\SU(N)$ with  $N_f=2N_c$. } \label{sun}
\end{center}
\end{figure}

In this section we briefly review the toric geometries that engineer $\mathcal{N}=2$  $\SU(N)$  theories. We first recall that the $\SU(N)$ theory with  $N_f=2N$ 
can be engineered  by the toric geometry in figure \ref{sun}.

The partition function  for this geometry (which agrees with the Nekrasov instanton partition) can be computed by gluing the left and right parts  (each of which is a strip) along the dotted line using the refined topological vertex (see e.g.~\cite{Taki:2007} for a discussion). 

Each of the two strips in the above geometry is related to special case of the $T_N$ theory.  This follows from the fact that in Gaiotto's language the $\SU(N)$ theory with  $N_f=2N$  theory is obtained via compactification on a  sphere  $C$ with four punctures (two basic $\U(1)$  punctures and two full  $\SU(N)$ punctures). In the weakly coupled degeneration limit where $C$ splits into two spheres, each sphere has one basic and two full punctures (one full puncture comes from the degeneration of the thin neck). Each sphere corresponds to a degenerate $T_N$ theory with one basic $\U(1)$ puncture and two full $\SU(N)$ punctures. We will refer to this theory as $\widetilde{T}_N$. Via the AGT conjecture, the $\tilde{T}_N$ theory is related to a (chiral) $A_{N-1}$ Toda  three-point function with one of the three primary fields of a special type \cite{Wyllard:2009}.

Let us give some details for the $\widetilde T_2$ case. The relevant toric strip diagram for  $\widetilde T_2$ takes the following form

\begin{figure}[h]
\begin{center}
\begin{pspicture}(7,5)
\psline[linecolor=red,linestyle=solid,linewidth=1pt]{-}(0,1)(6,1)
\psline[linecolor=red,linestyle=solid,linewidth=1pt]{-}(0,1)(0,4)
\psline[linecolor=red,linestyle=solid,linewidth=1pt]{-}(0,4)(6,4)
\psline[linecolor=red,linestyle=solid,linewidth=1pt]{-}(6,4)(6,1)
\psline[linecolor=red,linestyle=solid,linewidth=1pt]{-}(3,1)(3,4)
\psline[linecolor=red,linestyle=solid,linewidth=1pt]{-}(0,1)(3,4)
\psline[linecolor=red,linestyle=solid,linewidth=1pt]{-}(3,1)(6,4)
\psline[linecolor=black,linestyle=solid,linewidth=2pt]{-}(2,2.5)(4,2.5)
\psline[linecolor=black,linestyle=solid,linewidth=2pt]{-}(2,2.5)(1.5,3)
\psline[linecolor=black,linestyle=solid,linewidth=2pt]{-}(4,2.5)(4.5,2)
\psline[linecolor=black,linestyle=solid,linewidth=2pt]{-}(1.5,3)(1.5,5)
\psline[linecolor=black,linestyle=solid,linewidth=2pt]{-}(1.5,3)(-1,3)
\psline[linecolor=black,linestyle=solid,linewidth=2pt]{-}(2,2.5)(2,0)
\psline[linecolor=black,linestyle=solid,linewidth=2pt]{-}(4,2.5)(4,5)
\psline[linecolor=black,linestyle=solid,linewidth=2pt]{-}(4.5,2)(4.5,0)
\psline[linecolor=black,linestyle=solid,linewidth=2pt]{-}(4.5,2)(7,2)

\put(3.1,2.1){$Q_{f}$}
\put(1.5,2.3){$Q_{1}$}
\put(4.5,2.2){$Q_{2}$}
\put(1.6,4,7){$$}

\end{pspicture}
\caption{The $\widetilde T_2$ strip.} \label{T2strip}
\end{center}
\end{figure}
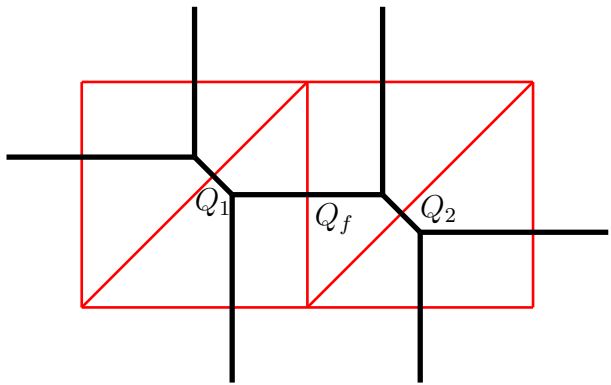

The partition function for this strip is
\bea\nonumber
\label{t2'}
 Z'_{\rm \widetilde T_2}=\prod_{i,j=1}^{\infty}\frac{(1{-}Q_{1}\,q^{-\rho_{i}}t^{-\rho_{j}})(1{-}Q_{f}\,q^{-\rho_{i}}t^{-\rho_{j}})(1{-}Q_{2}\,q^{-\rho_{i}}t^{-\rho_{j}})(1{-}Q_{1}Q_{f}Q_{2}\,q^{-\rho_{i}}t^{-\rho_{j}})}{(1-Q_{1}Q_{f}\,q^{-\rho_{i}+1/2}t^{-\rho_{j}-1/2})(1-Q_{f}Q_{2}\,q^{-\rho_{i}-1/2}t^{-\rho_{j}+1/2})}.\\
\eea
Clearly this partition function for the $\widetilde T_2$ geometry does not agree with the one for the $T_2$ geometry (\ref{t2})  (the closed topological vertex), even though in this case all three punctures are of the same type (basic). The latter geometry treats all punctures on an equal footing and has a manifest $\SU(2)^3$ flavour symmetry. However if we identify $Q_f=Q_1$ and use the dictionary (\ref{dict}) the partition functions  (\ref{t2}) and (\ref{t2'}) lead to four-dimensional expressions that  differ  by a product of  functions, each dependning only on one of the $\alpha_i$'s.  
This is not in contradiction with the AGT conjecture since in the Liouville theory we have the freedom to rescale each vertex operator by an arbitrary function of its momentum. The AGT conjecture is not sensitive to such functions (called $f(\alpha)$ in \cite{Alday:2009}) and in this sense the $\widetilde{T}_2$ strip also agrees with the (chiral) three-point function in the Liouville theory (cf.~also the discussions in the previous subsection and in \cite{Schiappa:2009}). 

The argument we gave for the $\widetilde T_2$ case  extends also  to the higher rank cases. The topological string partition function for the $\widetilde T_N$ strip geometry is consistent (in the above sense) with a  chiral version of the three-point function (with one vertex operator of special type) in the $A_{N-1}$ Toda theory.

In the next section we will compute open amplitudes (amplitudes with brane insertions) for the $\widetilde T_N$ geometries. It would also be interesting to compute open amplitudes for the $T_2$ geometry.

\subsection{Surface operators as toric branes}\label{torsurf}

In this subsection, we illustrate the claim that the gauge theory partition function in the presence of 
 a surface operator can be computed by an A-model topological string amplitude
with the insertion of a toric brane \cite{Gukov:2009}. In particular, based on results obtained in sections \ref{sCFT} and \ref{sSurf}, we arrive at the conclusion  that the insertion of a single surface operator corresponds to the insertion of a single toric brane, i.e.~the allowed representations on the external leg are given by Young tableaux in the form of columns. 

\begin{figure}[h]
\begin{center}
\begin{pspicture}(7,5)
\psline[linecolor=red,linestyle=solid,linewidth=1pt]{-}(0,1)(6,1)
\psline[linecolor=red,linestyle=solid,linewidth=1pt]{-}(0,1)(0,4)
\psline[linecolor=red,linestyle=solid,linewidth=1pt]{-}(0,4)(6,4)
\psline[linecolor=red,linestyle=solid,linewidth=1pt]{-}(6,4)(6,1)
\psline[linecolor=red,linestyle=solid,linewidth=1pt]{-}(3,1)(3,4)
\psline[linecolor=red,linestyle=solid,linewidth=1pt]{-}(0,1)(3,4)
\psline[linecolor=red,linestyle=solid,linewidth=1pt]{-}(3,1)(6,4)
\psline[linecolor=black,linestyle=solid,linewidth=2pt]{-}(2,2.5)(4,2.5)
\psline[linecolor=black,linestyle=solid,linewidth=2pt]{-}(2,2.5)(1.5,3)
\psline[linecolor=black,linestyle=solid,linewidth=2pt]{-}(4,2.5)(4.5,2)
\psline[linecolor=black,linestyle=solid,linewidth=2pt]{-}(1.5,3)(1.5,5)
\psline[linecolor=black,linestyle=solid,linewidth=2pt]{-}(1.5,3)(-1,3)
\psline[linecolor=black,linestyle=solid,linewidth=2pt]{-}(2,2.5)(2,0)
\psline[linecolor=black,linestyle=solid,linewidth=2pt]{-}(4,2.5)(4,5)
\psline[linecolor=black,linestyle=solid,linewidth=2pt]{-}(4.5,2)(4.5,0)
\psline[linecolor=black,linestyle=solid,linewidth=2pt]{-}(4.5,2)(7,2)

\put(3.1,2.1){$Q_{f}$}
\put(1.5,2.3){$Q_{1}$}
\put(4.5,2.2){$Q_{2}$}
\put(1.6,4,7){$\alpha$}

\end{pspicture}
\caption{Surface operator in $\widetilde T_2$} \label{st2}
\end{center}
\end{figure}
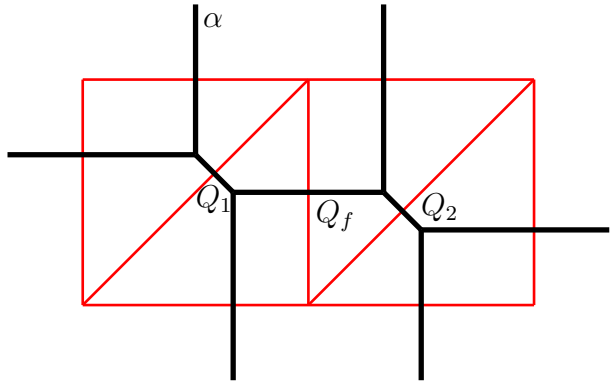

We  will focus on the  $\widetilde T_2$ geometry depicted in figure \ref{st2}. 
For this case it is possible to obtain closed expressions for the partition function 
in the presence of a surface operator also by using other methods, which provides us with non-trivial checks of our result.

The (unrefined $\cQ=0$) strip computation for the configuration in figure \ref{st2} leads to the following open amplitude (cf.~appendix \ref{striprules}):
\bea\label{strip}
Z_{\alpha}=s_{\alpha}(q^\rho)
\prod_{k}
\frac{
\left(1-q^k Q_1\right)^{C_k(\alpha,\bullet )}
\left(1-q^k Q_1 Q_2 Q_f\right)^{C_k(\alpha,\bullet)}}
{
\left(1-q^k Q_1 Q_f\right)^{C_k(\alpha,\bullet)}}\,.
\eea
We are interested in the special case of this expression which corresponds to the insertion of a single toric brane, which means that  the representation $\alpha$ should be a column of length $n$.  In this case we have:
\begin{equation}
C_{k}(\alpha,\bullet)=\left\{ 
\begin{array}{cc}
1 & \mbox{for}\, k\leq n\\
0 & \mbox{otherwise}
\end{array}
\right. \end{equation}
and the partition function in the sector corresponding to a column of length $n$ becomes
\be
Z_{(n)}(Q_1,Q_2,Q_f,q)=
\prod_{k=1}^n
\frac{
\left(1-q^k Q_1\right)
\left(1-q^k Q_1 Q_f Q_2\right)}
{\left(1-q^k \right)
\left(1-q^k Q_1 Q_f\right)
}\,.
\ee
We can now package the contributions from all columns into the total partition function:
\be
\label{open}
Z_{\rm open}(z)=\sum_{n=0}^{\infty} z^n Z_{(n)}(Q_1,Q_2,Q_f,q)\,.
\ee
With the dictionary (\ref{dict}), and in the four dimensional limit, 
this partition function agrees with $Z_{\rm null}(x)\Big |_{\cQ=0}$ given in (\ref{2F1}). 
 It also agrees with the result (\ref{2dinst}) obtained in the next section by combining the conjectures in \cite{Alday:2009} and \cite{Alday:2009b}. Furthermore, (\ref{open}) agrees with the conjectural $q$-deformed matrix model result discussed in \cite{Schiappa:2009}.
By putting arbitrary non-trivial  representations also on the two lower legs in figure \ref{st2} one can use the resulting diagram as a building block to obtain an expression for the partition function of   the $\SU(2)$ with $N_f{=}4$ theory with a surface operator insertion, that can be compared to the result obtained using the method described in the next section. 

Above we matched an A-model open generating function to a B-model computation, in the same spirit as in  \cite{Bouchard:2007}. However, there is an important difference: in \cite{Bouchard:2007}, the B-model open amplitudes were matched to A-model generating functions with the open moduli being holonomies $\mbox{tr}_{\lambda}\,V$ in all possible representations $\lambda$. This has the interpretation that the boundary conditions were provided by a stack of infinitely many toric branes which can carry arbitrary representations. In the present case, the A-model amplitude receives contributions only from representations with a single column since we only have a single toric brane insertion, which dovetails nicely with the fact that it corresponds to a single surface operator insertion on the gauge theory side. We will provide more evidence supporting this claim in what follows by showing that the number of surface operator insertions is equal to the number of columns in the representations.  Furthermore, it is interesting to note  that the M-theory and the engineering differentials appear to distinguish these two cases, i.e.~the first one computes amplitudes with one-column representations while the second is relevant for stacks of branes. Although the integrals over closed cycles are not sensitive to different choices of Seiberg-Witten differentials, open integrals clearly are.

\par{Let us work out the open topological amplitude corresponding to multiple surface operator insertions. We need to put a stack of branes on the external leg consisting of as many branes as the number of surface operators inserted, say $m$ branes. In this case, we need to consider representations with $m$ columns and figure out the exponents $C_{k}(\alpha,\bullet)$ for such a representation. If we look at the definition of these exponents we realize that they can be written in a more suggestive form:  }

\bea
\sum_{k}C_{k}(\alpha,\bullet)q^{k}=q\sum_{i=1}^{d_{\alpha}}q^{-i}\sum_{j=0}^{\alpha_{i}-1}q^{j}=\sum_{i=1}^{d_{\alpha}}\sum_{j=1-i}^{\alpha_{i}-i}q^{j}.
\eea
This implies that for the $i^{\rm th}$ column the exponents are 1 for $1-i\leq k \leq\alpha_{i}-i$, and 0 otherwise. Hence, for the $i^{\rm th}$ column we have
\bea\nonumber
\left .\prod_{k}\left(1-q^{k}Q\right)^{C_{k}(\alpha,\bullet)}\right|_{\alpha_{i}}&=&\prod_{k=1-i}^{\alpha_{i}-i} \left(1-q^{k}Q\right)=\prod_{k=0}^{\alpha_{i}-1} \left(1-q^{k-i+1}Q\right)\\
&=&\left(Q\, q^{1-i};q\right)_{\alpha_{i}},
\eea
where $(a;q)_{k}$ is known as the $q$-shifted factorial. The factors appearing in the open amplitude can then be written as:
\bea\label{factorial}
\prod_{k}\left(1-q^{k}Q\right)^{C_{k}(\alpha,\bullet)}=\prod_{i=1}^{d_{\alpha}}\left(Q\, q^{1-i};q\right)_{\alpha_{i}}.
\eea
The above form of the open topological string amplitude makes it easier to compare with the Liouville result. However, we need to keep in mind that the Liouville theory provides a result which corresponds to a $4d$ gauge theory expression whereas the topological vertex computation gives the $5d$ gauge theory result before taking the field theory limit. It is known in many cases that the $4d$ and $5d$ results are related by the so-called $q$-deformation. Here we will show that the vertex result is a certain $q$-deformation of the generalized hypergeometric function which appears in the $4d$ theory, cf.~(\ref{hyperJack}). The hypergeometric function obeys a differential equation and its $q$-deformation a difference equation\footnote{The possibility that normalized open string amplitudes obeys difference equations was suggested to one of us by D. Gaiotto.}. The $qt$-deformed hypergeometric function \cite{Kaneko:1996}, is defined as
\bea
\label{hyperfunc}
{}_{r}\Phi_{s}^{(q,t)}(a_{1},\mathellipsis,a_{r};b_{1},\mathellipsis,b_{s};\mathbf{z}) \sum_{\lambda}\frac{\prod_{i=1}^{r}\left(a_{i}\right)_{\lambda}^{(q,t)}\left \{(-1)^{|\lambda|}q^{n(\lambda^{t})} \right\}^{s+1-r}}{\prod_{i=1}^{s}\left(b_{i}\right)_{\lambda}^{(q,t)}h'_{\lambda}(q,t)}P_{\lambda}(\mathbf{z};q,t).
\eea
The notation in this definition requires some clarification: $P_{\lambda}(\mathbf{z};q,t)$ is the Macdonald polynomial (function), with $\mathbf{z}=(z_1,\ldots,z_k)$, $\left(a\right)_{\lambda}^{(q,t)}$ is defined in terms of the arm-colength $a'(s)=j-1$ and leg-colength $\ell'(s)=i-1$ of the Young diagram $\lambda$, and
\bea
\left (a\right)_{\lambda}^{(q,t)}=\prod_{s\in\lambda}\left (t^{\ell'(s)}-q^{a'(s)}\,a \right).
\eea
The function $h'_{\lambda}(q,t)$ is defined as a product of factors including the arm-length $a(s)=\lambda_{i}-j$ and the leg-length $\ell(s)=\lambda^t_{j}-i$ 
\bea
h'_{\lambda}(q,t)=\prod_{s\in\lambda}\left (1-q^{a(s)+1}t^{\ell(s)} \right).
\eea
Lastly, $n(\lambda)=\sum (i-1)\lambda_{i}$. 

We now show that the $qt$-deformed hypergeometric function is related to the strip computation when we set $t=q$ and manipulate the factors:
\bea
\left(a\right)^{(q,q)}_{\lambda}=q^{n(\lambda)}\prod_{i=1}^{d_{\lambda}}\prod_{j=1}^{\lambda_{i}}(1-a\,q^{j-i})=q^{n(\lambda)}\prod_{i=1}^{d_{\lambda}}\left (a\,q^{1-i};q \right)_{\lambda_{i}},
\eea
where we have used the identity $n(\lambda)=\sum_{s\in\lambda}\ell'(s)$.
By using (\ref{factorial}) we get
\bea
\left(a \right)^{(q,q)}_{\lambda}=q^{n(\lambda)}\prod_{k}\left(1-a\,q^{k}\right)^{C_{k}(\lambda,\bullet)}.
\eea
The form of the strip result imposes $s=1$ and $r=2$, hence, $s+1-r=0$. We have three factors of $q^{n(\lambda)}$, two in the numerator and one on the denominator. The remaining factor can be combined with $h'_{\lambda}(q,q)$ in the denominator to give the Schur function:
\bea\nonumber
s_{\lambda}(1,q,q^{2},\mathellipsis)&=&q^{n(\lambda)}\prod_{s\in\lambda} \frac{1}{1-q^{a(s)+\ell(s)+1}}\\
&=&q^{-\frac{|\lambda|}{2}}s_{\lambda}(q^{-\rho})\, .
\eea 
Having reproduced also the Schur function as the pre-factor in the strip computation we find  agreement between the vertex computation and the $qt$-deformed hypergeometric function. The factor $q^{-\frac{|\lambda|}{2}}$ can be absorbed in the Macdonald function and the definition of the the open moduli $\mathbf{z}$.   

\begin{figure}[h]
\begin{center}
\begin{pspicture}(12,5)

\psline[linecolor=red,linestyle=solid,linewidth=1pt]{-}(0,1)(6,1)
\psline[linecolor=red,linestyle=dotted,linewidth=1pt]{-}(6,1)(7,1)
\psline[linecolor=red,linestyle=solid,linewidth=1pt]{-}(0,1)(0,4)
\psline[linecolor=red,linestyle=solid,linewidth=1pt]{-}(0,4)(6,4)
\psline[linecolor=red,linestyle=dotted,linewidth=1pt]{-}(6,4)(7,4)
\psline[linecolor=red,linestyle=solid,linewidth=1pt]{-}(6,4)(6,1)
\psline[linecolor=red,linestyle=solid,linewidth=1pt]{-}(3,1)(3,4)
\psline[linecolor=red,linestyle=solid,linewidth=1pt]{-}(0,1)(3,4)
\psline[linecolor=red,linestyle=solid,linewidth=1pt]{-}(3,1)(6,4)
\psline[linecolor=black,linestyle=solid,linewidth=2pt]{-}(2,2.5)(4,2.5)
\psline[linecolor=black,linestyle=solid,linewidth=2pt]{-}(2,2.5)(1.5,3)
\psline[linecolor=black,linestyle=solid,linewidth=2pt]{-}(4,2.5)(4.5,2)
\psline[linecolor=black,linestyle=solid,linewidth=2pt]{-}(1.5,3)(1.5,5)
\psline[linecolor=black,linestyle=solid,linewidth=2pt]{-}(1.5,3)(-1,3)
\psline[linecolor=black,linestyle=solid,linewidth=2pt]{-}(2,2.5)(2,0)
\psline[linecolor=black,linestyle=solid,linewidth=2pt]{-}(4,2.5)(4,5)
\psline[linecolor=black,linestyle=solid,linewidth=2pt]{-}(4.5,2)(4.5,0)
\psline[linecolor=black,linestyle=solid,linewidth=2pt]{-}(4.5,2)(6,2)
\psline[linecolor=black,linestyle=dotted,linewidth=2pt]{-}(6,2)(7,2)

\psline[linecolor=red,linestyle=solid,linewidth=1pt]{-}(9,1)(12,1)
\psline[linecolor=red,linestyle=dotted,linewidth=1pt]{-}(8,1)(9,1)
\psline[linecolor=red,linestyle=solid,linewidth=1pt]{-}(9,1)(9,4)
\psline[linecolor=red,linestyle=solid,linewidth=1pt]{-}(9,4)(12,4)
\psline[linecolor=red,linestyle=solid,linewidth=1pt]{-}(12,4)(12,1)
\psline[linecolor=red,linestyle=solid,linewidth=1pt]{-}(9,1)(12,4)
\psline[linecolor=red,linestyle=dotted,linewidth=1pt]{-}(8,4)(9,4)
\psline[linecolor=black,linestyle=dotted,linewidth=2pt]{-}(8,2)(9,2)
\psline[linecolor=black,linestyle=solid,linewidth=2pt]{-}(9,2)(9.5,2)
\psline[linecolor=black,linestyle=solid,linewidth=2pt]{-}(9.5,2)(10,1.5)
\psline[linecolor=black,linestyle=solid,linewidth=2pt]{-}(10,1.5)(10,0)
\psline[linecolor=black,linestyle=solid,linewidth=2pt]{-}(10,1.5)(13,1.5)
\psline[linecolor=black,linestyle=solid,linewidth=2pt]{-}(9.5,2)(9.5,5)

\put(3.1,2.1){$Q_{f,1}$}
\put(1.5,2.3){$Q_{1}$}
\put(4.5,2.2){$Q_{2}$}
\put(1.6,4,7){$\alpha$}
\put(10,1.7){$Q_{N}$}

\end{pspicture}
\caption{Surface operator in $\widetilde{T}_N$} \label{stnt}
\end{center}
\end{figure}
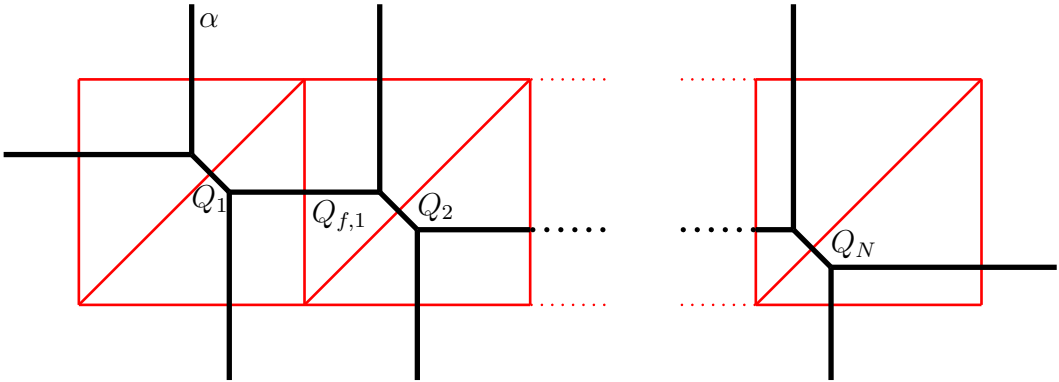

We can further generalize our proof of the equivalence to the $\SU(N)$ gauge theory with $2N$ hypermultiplets in the fundamental representation.  Half of the geometry engineering this theory, in our language $\widetilde{T}_{N}$, is depicted in figure \ref{stnt}. 

This theory was studied in \cite{Iqbal:2004} and the normalized result of the strip computation with all external legs labeled non-trivially (the upper legs by $\{\alpha_{i}\}$ and lower ones by $\{\beta_{i} \}$) is given by: 
\bea 
\nonumber
K^{\alpha_{1}\cdots \alpha_{N}}_{\beta_{1}\cdots\beta_{N}} &=&\frac{{\cal 
K}^{\alpha_{1}\cdots \alpha_{N}}_{\beta_{1}\cdots\beta_{N}}}{ 
{\cal 
K}^{\bullet \cdots \bullet}_{\bullet \cdots \bullet}} =\label{basic} s_{\alpha_{1}}s_{\beta_{1}}\cdots s_{\alpha_{N}}s_{\beta_{N}}\,
\\
&\times &\prod_{k}\frac{\prod_{i\leq 
j}(1-q^{k}Q_{\alpha_{i}\beta_{j}})^{C_{k}(\alpha_{i},\beta_{j})} 
\prod_{i<j}(1-q^{k}Q_{\beta_{i}\alpha_{j}})^{C_{k}(\beta_{i}^t,\alpha_{j}^{t})}} 
{\prod_{i<j}(1-q^{k}Q_{\alpha_{i}\alpha_{j}})^{C_{k}(\alpha_{i},\alpha_{j}^{t})} 
(1-q^{k}Q_{\beta_{i}\beta_{j}})^{C_{k}(\beta_{i}^t,\beta_{j})}}, 
\eea 
where the K\"ahler parameters $Q_{\alpha,\beta}$ are: 
\bea Q_{\alpha_{i}\alpha_{j}}&=&Q_{ij}\\\nonumber 
Q_{\alpha_{i}\beta_{j}}&=&Q_{ij}Q_{j}\\\nonumber 
Q_{\beta_{i}\alpha_{j}}&=&Q_{ij}Q_{i}^{-1}\,,\\\nonumber 
Q_{\beta_{i}\beta_{j}}&=&Q_{ij}Q_{i}^{-1}Q_{j}\,, \eea with
$Q_{ij}=\prod_{k=i}^{j-1}Q_{k}Q_{f,k}$.

When we insert  $m$ branes on one of the external legs this  corresponds, in the strip language, to putting all the representations trivial except one, say $\alpha_{1}=\alpha$. 
As before we can show that the coefficients $C_{k}(\alpha,\bullet)$ match
the $q$-shifted factorials in the definition of the $q$-deformed hypergeometric function. Notice that there are $r=N$ factors in the numerator and $s=N-1$ in the denominator. 
Finally the prefactor $s_{\alpha}$ can be reproduced following the argument given above.

We close this section by sketching  how to extend the  above results  to the refined case. The starting point is the refined strip partition function for the $\widetilde T_N$ geometry~\cite{Taki:2007}:
\begin{align}\label{refopen}\nonumber
&K_
{\beta_{1}\beta_{2}\mathellipsis}^{\alpha_{1}\alpha_{2}\mathellipsis}=
\prod_{a}\left [q^{\frac{\Arrowvert\alpha_{a}\Arrowvert^{2}}
{2}}t^{\frac{\Arrowvert\beta{a}\Arrowvert^{2}}{2}} 
\widetilde{Z}_{\alpha_{a}}(t,q)\widetilde{Z}_{\beta_{a}(q,t)}\right]\\ 
\nonumber
&\times\prod_{i,j=1}^{\infty}\prod_{1\leq a\leq b\leq N}\left(1{-}
Q_{\alpha_{a}\beta_{b}}\,t^{-\alpha_{a,i}^{t}{+}j{-}\frac{1}{2}}q^{-\beta_{b,j}^{t}
{+}i{-}\frac{1}{2}} \right)\!\! \prod_{1\leq a < b\leq N} \!\! \left(1{-}Q_{\beta_{a}\alpha_{b}}
\, t^{-\beta_{a,i}{+}j{-}\frac{1}{2}}q^{-\alpha_{b,j}{+}i{-}\frac{1}{2}} \right )\\
&\times\prod_{1\leq a < b\leq N}\left(1-Q_{\alpha_{a}\alpha_{b}}\, t^{-
\alpha^{t}_{a,i}+j}q^{-\alpha_{b,j}+i-1} \right )^{-1}\left (1-
Q_{\beta_{a}\beta_{b}}\,t^{-\beta_{a,i}+j-1}q^{-\beta_{b,j}^{t}+i} 
\right)^{-1},
\end{align}
where $\widetilde{Z}_{\nu}(t,q)=\prod_{s\in \nu}(1-t^{a(s)+1}  q^{\ell(s)})^{-1}.$
As above, we  set all  the representations trivial except one, say $\beta_{N}=\beta$. 
To match to the $qt$-deformed hypergeometric function (\ref{hyperfunc}), we need to normalise the  above amplitude by the closed one; in this way 
we are left with  $N$ factors in the numerator and $N-1$ in the 
denominator, as in the unrefined case.

Let us focus on a particular factor in the numerator normalized by the corresponding closed one. Taking the logarithm, expanding and using the following identities:
\bea\nonumber
\sum_{i,j=1}^{\infty}\left (t^{j-1}q^{-\beta_{j}^{t}+i}-t^{j-1}q^{i}
\right)&=&\sum_{j=1}^{d_{\beta^{t}}}t^{j-1}\sum_{i=1}^{\infty}
\left(q^{-\beta_{j}^{t}+i}-q^{i}\right)=\\
&=&\sum_{j=1}^{d_{\beta^{t}}}\sum_{i=1}^{\beta_{j}^{t}}t^{j-1}q^{-
(i-1)}=\sum_{s\in\beta^{t}}t^{a'(s)}q^{-\ell'(s)},
\eea
we  get:
\bea\label{prefactor}
\frac{\prod_{i,j=1}^{\infty}\left(1-Q\,t^{j-1/2}q^{-\beta_{j}^{t}
+i-1/2} \right)}{\prod_{i,j=1}^{\infty}\left(1-Q\,t^{j-1/2}q^{i-1/2} 
\right)}=q^{-n(\beta^{t})}(\widetilde{Q}  )^{(t,q)}_{\beta^{t}} \,,
\eea
where $\widetilde{Q}=Q\,t^{1/2}q^{-1/2}$.
Notice we have only one  factor of $q^{-n(\beta^{t})}$ which combines with  $t^{\frac{\Arrowvert\beta{a}\Arrowvert^{2}}{2}} $ to give a refined framing factor, up to factors that can be absorbed in the open moduli. Finally  the factor  $h'_{\beta}(q,t)$ in the denominator of eq. (\ref{hyperfunc})   matches $\widetilde{Z}_{\beta}(q,t)$ in (\ref{refopen}).

\setcounter{equation}{0} 
\section{AGT approach: Gauge theory and instanton counting}  \label{sSurf}

In the previous section we made a gave an explicit prescription for how to determine the instanton partition function for an $\SU(N)$ quiver gauge theory in the presence of a surface operator using the conjectural relation \cite{Gukov:2009} to an A-model topological string partition function with toric brane insertions. 
To obtain similar closed expressions directly in gauge theory would require extending the instanton counting method to gauge theories with surface operator insertions. However, as we shall see, even without the full machinery some insights can be gained by using the conjectures in \cite{Alday:2009,Alday:2009b}. We start by recalling some pertinent facts about the instanton counting method (for further details, see e.g.~\cite{Nekrasov:2002,Bruzzo:2002}).

The Nekrasov partition function (from which the prepotential and the Seiberg-Witten curve can be obtained) factorises into two parts as  
\be \label{z}
Z = Z_{\rm pert}\,Z_{\rm inst}\,,
\ee 
where $Z_{\rm pert}$ is the contribution from perturbative calculations (there are contributions only at tree and one--loop level), and $Z_{\rm inst}$ is the contribution from instantons. 

The instanton partition function for  an arbitrary $A_r$ quiver theory with matter multiplets in bifundamental and fundamental representations can be written down in closed form using 
 various building blocks that involve partitions (or equivalently Young tableaux) in a fundamental way.  
 
The Coulomb branch of an $A_r$ quiver gauge theory is parameterised by the Coulomb branch moduli $\ha^I_n$, where $n=1,\ldots N_I$ and $I$ label the various nodes of the quiver. Since we are interested in $\SU(N_I)$ gauge groups rather than $\U(N_I)$ we impose the restrictions $\sum_{n=1}^{N_I} \ha^I_n =0$. In the particular case of $\SU(2)$ this translates into $\ha = (a,-a)$. 
For each  $\ha^I_m$  Coulomb modulus (i.e.~before restricting to $\SU(N_I)$) there is an associated Young tableau; e.g.~for $\SU(2)$ one has a pair of Young tableaux. 

It is convenient to define 
\be
E(x,Y^I_n,Y^J_m,s_I) = x-\ep_1 L_{Y^{J}_m}(s) + \ep_2(A_{Y^I_n}(s)+1)\,,
\ee
 where $s_I=(i,j)$ and $i$ refers to the vertical position and $j$ to the horizontal position of a box in the Young tableau $Y^I_n$. Furthermore, $L_{Y^J_m} = k_{m,i}-j$ and $A_{Y^I_n} = k^T_{n,j} - i$, where $k_{m,i}$ is the length of the $i$th row of $Y^J_m$ and $k^T_{n,j}$ is the height of the $j$th column of $Y^I_n$.

 For a bifundamental matter multiplet  of mass $m$ connecting node $I$ with node $J$ one gets the contribution
\be
[E(a^I_n-a^J_m,Y^I_n,Y^J_m,s_I)-m] [ \ep - E(a_m^J-a^I_n,Y^J_m,Y^I_n,s_J)-m  ]\,,
\ee
where  $\ep \equiv \ep_1+\ep_2$. The gauge multiplet of node $I$ contributes
 \be
\frac{1}{ E(a^I_n-a_m^I,Y^I_n,Y^I_m,s) [ \ep - E(a^I_n-a_m^I,Y^I_n,Y^I_m,s) ] }\,,
 \ee
 and a matter field of mass $m$ transforming in the fundamental representation of gauge group $I$ contributes  a factor  
\be
P(a_n^I,Y_n,s,m) = a_n^I+(j-1)\ep_1+(i-1)\ep_2 - m\,.
\ee

As an example, the instanton partition function for the $\SU(N)\times\SU(N)$ theory with one bifundamental hypermultiplet and $N$ fundamentals at each of the two nodes can be 
written 
\bea \label{Z2quiver}
&& \sum_{\vec{Y},\vec{W}} y^{|\vec{Y}|} z^{|\vec{W}|}  \prod_{n,m=1}^{N} \prod_{s\in Y_n} \frac{ [E(\ha^1_n-\ha^2_m,Y_n,W_m,s)-m]\prod_{f=1}^{N}P(\ha^1_n,Y_n,s,m_f) }{E(\ha^1_n-\ha^1_m,Y_n,Y_m,s)[E(\ha^1_n-\ha^1_m,Y_n,Y_m,s) -\ep] } \non \\ 
&& \times \prod_{t\in W_n} \frac{[\ep-E(\ha^2_n-\ha^1_m,W_n,Y_m,t)-m]\prod_{f=1}^{N}
P(\ha^2_n,W_n,t,\tilde{m}_f) }{E(\ha^2_n-\ha^2_m,W_n,W_m,t)[E(\ha^2_n-\ha^2_m,W_n,W_m,t) -\ep] } 
\,,
\eea
where, for the first node, $\vec{Y}$ denotes the  $N$-dimensional vector of Young tableaux, $(Y_1,Y_2,\ldots
,Y_{N})$, $\vec{a}^1=(\ha^1_1,\ldots,\ha^1_N)$ are the corresponding moduli, and $y=\re^{2\pi {\rm i} \tau_1}$, where $\tau_1$ is the complexified gauge coupling constant, and $|\vec{Y}|$ (the instanton number) is the total number of boxes in all the $Y_n$'s. The objects $\vec{W}$, $\ha^2_n$, and $z= \re^{2\pi {\rm i} \tau_2} $ denote the corresponding quantities for the second node. 

The AGT relation relates the instanton partition function of an $A_r$ quiver theory to a certain chiral block in the $A_r$ Toda theory. For instance, the above expression (\ref{Z2quiver}) should be related to a five-point chiral block. For $\SU(2)$ quivers, multi-point conformal blocks have been matched to the corresponding instanton partition functions in~\cite{Alday:2009,Alba:2009}. 

The extension of the AGT conjecture presented in \cite{Alday:2009b} states that certain degenerate vertex operator insertions should correspond to surface operator insertions on the gauge theory side. Consider the prototypical case of inserting a single $V_{-b/2}$ operator into the four-point function in the Liouville theory. In the perturbative approach, using the degenerate fusion rules, the resulting expression can be viewed as a restriction of a generic five-point function; schematically
\be \label{5ptrelation}
\lb \al_1| V_{\al_2} |\si\rb \lb \si|  V_{\al_3} |\al_4{+}{\ts \frac{b}{2}} \rb \lb \al_4 {+}{\ts \frac{b}{2}} | V_{-\frac{b}{2} } |\al_4 \rb =  
\lb \al_1| V_{\al_2} |\si \rb \lb \si |  V_{\al} |\tilde{\si} \rb \lb \tilde{\si} | V_{\al_3} |\al_4 \rb \bigg|_{ \stackrel{\scriptstyle  \tilde{\si} =\al_4+b/2  }{ \! \! \! \al_3=-b/2 }  }  
\ee
Now, via the AGT conjecture, a five point function can be related to the instanton partition function in the $\SU(2){\times}\SU(2)$ theory with a bifundamental matter multiplet of mass $m$ and two fundamental matter multiplets in each of the two $\SU(2)$ factors. This is simply the above expression (\ref{Z2quiver}) restricted to $N=2$. We write $\vec{\ha}^1=(a,-a)$ and $\vec{\ha}^2=(\tila,-\tila)$. Furthermore, using the AGT relation (in our conventions)
\be
\ba{rclrclrcl}
  -\al_1 - \al_2 &=& m_1 - \frac{\ep}{2} \,, &&&& \, -\al_4 - \al_3&=&\tilde{m}_1 - \frac{\ep}{2} \,,  
  \\ [3pt] 
 \al_1- \al_2&=&  m_2  + \frac{\ep}{2} \,,   &&&& \al_4 -\al_3 &=& \tilde{m}_2 + \frac{\ep}{2} \,, 
  \\ [3pt]  
  \si &=& a+\frac{\ep}{2} \,, & \al &=& m \,, &  \tilde{\si} &=& \tila +\frac{\ep}{2}\,,  
\ea
\ee
the restrictions in (\ref{5ptrelation}) can also be translated into gauge theory language:
\be
\tila =  \tilde{m}_2  = - \tilde{m}_1 + \ep_1  \,.
\ee
Imposing these restrictions allows us to simplify the expression. Since $\tila=\tilde{m}_2$ only terms in the sum with only $W_2$ non--empty give non--vanishing contributions \cite{Mironov:2009b}. If we furthermore set $\tilde{m}_1= - \tila + \ep_1$ then only those $W_2$ tableaux that have boxes only in the first column survive \cite{Mironov:2009b} (in other words $k_{2,j}^T$ is only non--zero for $j=1$, so that $i=1,\ldots,k_{2,1}^T$ and $k_{2,i}=1$). Using this result we find
\bea \label{SU2surf}
&& \sum_{\ell=0}^\infty \sum_{\vec{Y}} y^{|\vec{Y}|} z^{\ell}  \prod_{n,m=1}^{2} \prod_{s\in Y_n} \frac{ [E(\ha^1_n,Y_n,W,s)-m_4-\ep] \prod_{f=1}^{3}P(\ha^1_n,Y_n,s,m_f) }{E(\ha^1_n-\ha^1_m,Y_n,Y_m,s)[E(\ha^1_n-\ha^1_m,Y_n,Y_m,s) -\ep] } \non \\ 
&& \times \prod_{t\in W}  \frac{[ - E(-\ha^1_m,W ,Y_m,t) - m_4] }{ [m_4-m_3+\ep + \ep_2\, (p-1)][ \ep_2 \, p] } 
\,.
\eea
where we have introduced the notation
\be
m_3=m+\tila-\ep \,, \qquad m_4=m-\tila-\ep \,.
\ee
(Note that the product over $f$ in the numerator of (\ref{SU2surf}) runs over three masses.) Furthermore, in (\ref{SU2surf})  $W$ denotes a Young tableaux with only one column, where $p=1,\ldots,\ell$ label the boxes. 

To recap, we started from a four point function/conformal block in the Liouville theory with an additional degenerate insertion and used the AGT conjecture to rewrite it in terms of a instanton partition function in an $\SU(2){\times}\SU(2)$ theory with additional restrictions, which could be further simplified to the expression (\ref{SU2surf}). 

But if the conjecture in~\cite{Alday:2009b} is correct  (\ref{SU2surf}) should correspond to the instanton partition function for the $\SU(2)$ gauge theory with four fundamental matter multiplets together with a surface operator insertion. The above expression (\ref{SU2surf}) has a form which agrees with  general expectations. It has a sum over conventional instantons labelled by a pair of Young tableaux as well as a sum over ``two-dimensional instantons" due to the surface operator, labelled by an integer $\ell$. Thus it seems very plausible that the above expression really represents the gauge theory partition function in the presence of a surface operator.

In particular, the terms with $\ell=0$ are easily seen to reproduce the usual instanton expansion, whereas for the terms with $|\vec{Y}|=0$ one finds
\be \label{2dinst}
Z_{\rm 2d \,\, inst} = \sum_{\ell=0}^{\infty} \frac{(A_1)_\ell(A_2)_{\ell}}{(B_1)_{\ell} }\frac{z^\ell}{\ell!} \,,
\ee
where  
\be \label{hejhopp}
A_1 = \frac{1}{\ep_2}(m_4+a+\ep) \,, \quad A_2=\frac{1}{\ep_2}(m_4-a+\ep)  \,, \quad B_1 = \frac{1}{\ep_2}(m_4-m_3+\ep)
\ee
and $(X)_n= X (X+1) \cdots (X+n-1)$ is the Pochhammer symbol. This is simply the series expansion of the hypergeometric function ${}_2 F_1(A_1,A_2;B_1;z)$. Translating to CFT language using $b=1/\ep_2$ and
\be
m_4=\al_3+\al_4-{\ts \frac{3}{2} }\ep \,, \quad m_3 = \al_3-\al_4-\half \ep \,, \quad a = -\si + \half \ep \,,
\ee
we find that (\ref{hejhopp}) becomes 
\be
A_1 = b\,(\al_4+\al_3-\si)  \,, \quad A_2= b\,(\al_4+\al_3 + \si-\ep)  \,, \quad B_1 = 2 \, b\,\al_4
\ee
The above expression is easily seen to agree with the result we obtained using the A-model topological string in the previous section as well as with the result in section~\ref{sCFT}.

In the usual  instanton counting method an alternative way to write the instanton expansion is in terms of  certain contour integrals together with a particular choice of contour, leading to a prescription for which poles are picked up by the integration. Such an expression is also possible  for the above result (\ref{2dinst}) which alternatively can be written as
\be \label{loco}
 \sum_{\ell=0}^{\infty} \frac{z^\ell}{\ep_2^\ell \ell!}   \oint \prod_{i=1}^\ell \D\, \xi_i \frac{(\xi_i + \mu_1) (\xi_i + \mu_2) }{ (\xi_i -\nu ) (\xi_i+\nu) } \prod_{1\le i < j \le \ell } \frac{(\xi_i-\xi_j)^2}{(\xi_i-\xi_j-\ep_2/2)(\xi_i-\xi_j+\ep_2/2)}   
\ee
where we used the notation
\be
\nu  = \half(-m_3+m_4)  \,, \quad \mu_1= -\half(m_3+m_4) + a - \ep \,, \quad \mu_2= -\half(m_3+m_4) - a - \ep \,,
\ee
and the poles with positive $\ep_2$ part in (\ref{loco}) are selected (if there is no $\ep_2$ part then the poles with positive $\nu$ part are selected). 
This expression can be viewed as arising from a localisation problem. See \cite{Nekrasov:2009} for a discussion of similar issues. The above integrals can also be viewed as computing the character of an (equivariant) instanton deformation complex. 

Above the degenerate state was inserted using a particular ordering of the vertex operators.  However, other orderings can also be treated using the above method. For instance, another way to introduce a degenerate state is via:
\be
\lb \al_1| V_{\al_2} |\si{-}{\ts \frac{b}{2}} \rb \lb \si{-}{\ts \frac{b}{2}}|  V_{-\frac{b}{2}} |\si \rb \lb \si | V_{\al_3} |\al_4 \rb =  
\lb \al_1| V_{\al_2} |\si' \rb \lb \si' |  V_{\al} |\si \rb \lb \si | V_{\al_3} |\al_4 \rb \bigg|_{ \stackrel{\scriptstyle \si' =\si-b/2}{\!\! \al=-b/2} }  
\ee
Again, the restrictions can be translated into gauge theory language and the resulting expression can be simplified. We expect the result to be related to the insertion of a toric brane on an internal line in the toric diagram. 

Using the above arguments it is clear that one can also obtain (conjectural) gauge theory expressions corresponding to  insertions of other degenerate states in the class $V_{-p \frac{b}{2}-q\frac{1}{2b}}$ as well as expressions for the higher rank $\SU(N)$ theories\footnote{The  conditions reducing the sums in the $\SU(N)$ Nekrasov partition function to a sum over only columns were discussed in \cite{Mironov:2009b,Mironov:2009}.}.

\setcounter{equation}{0} 
\section{Summary and outlook} \label{sdisc}

In this paper we studied surface operators and their conjectured interpretation as degenerate operators in the dual Toda field theories. 

In particular, we analysed the surface operators using methods from topological string theory. We showed that by making use of  the B-model topological recursion method one can compute the partition function including the effects of (one or many) surface operators beyond the semi-classical limit order-by-order. 

An alternative viewpoint is the interpretation of the surface operator as a toric brane in the A-model topological string \cite{Gukov:2009}. This identification extends the usual relation between the topological string partition function and the Nekrasov instanton partition function, and makes algorithmic computations possible using the (refined) topological vertex. 

There are several remaining questions and possible extensions of our work. Here we briefly mention some of them.

$\bullet$ The gauge theory instanton counting method in the presence of surface operators should be  developed. Our results in sections \ref{sSurf} and \ref{sA} should provide important clues. The brane realisation of a surface operator \cite{Constable:2002,Gukov:2006} may be a useful tool.

$\bullet$ The classification of surface operators and the precise relation to the classification of degenerate operators and toric branes should be clarified and developed in full generality. 

$\bullet$ The topological recursion method was only used for the case $\ep=0$. It would be desirable to extend it also to the $\ep\neq 0$ setting (beta deformed case). Is there a way to understand the recursion method directly in the CFT?\footnote{For a recent discusion of the topological recursion in a CFT context see \cite{Kostov:2009}. }

$\bullet$ The toric approach to the $T_N$ theories may provide insights into the important problem of calculating general three-point functions in the $A_{N-1}$ Toda field theories. 

$\bullet$ The B-model approach has the advantage that (using mirror symmetry techniques) it can be used in all regions of the moduli space. It should therefore be a useful tool to address strong coupling questions such as  S-duality and crossing symmetry.

$\bullet$  It seems natural to view the $T_N$ diagrams as composite vertices. These vertices can be glued\footnote{The $T_2$ model can alternatively be described in terms of a certain crystal model \cite{Sulkowski:2006} and it was speculated in \cite{Sulkowski:2006} that it might be possible to use it as a building block for more complicated geometries.}  and lead to a sort of fattening of the diagrams drawn in \cite{Gaiotto:2009}. What do the resulting toric diagrams describe and are they useful? 
\section*{Acknowledgements}

We would like to thank P. Berglund, A. Brini, V. Fateev, W. Lerche, M. Mari\~no, N. Orantin, F. Passerini, S. Ribault, R. Schiappa, S. Shatashvili, Y. Tachikawa and J. Walcher for discussions. 
We would like to thank S. Gukov for bringing the ASC, Munich talk listed in \cite{Gukov:2009} to our attention. 

NW would like to thank CERN PH-TH for generous hospitality and financial support during the course of this work. He would also like to thank the Yukawa Institute for generous hospitality and financial support as well as the organisers and participants of the inspiring workshop `Recent advancements in gauge theories and CFTs'.


\appendix

\setcounter{equation}{0} 
\section{Appendix}

\subsection{Jack polynomials}\label{jack}
The Jack polynomials are homogeneous polynomials of $k$ variables that depend on a continuous parameter $\beta=-\frac{1}{b^2}$, and are indexed by partitions. They are eigenfunctions of the hamiltonian
\be
\frac{1}{2}\sum_i x_i^2 \frac{\pa^2}{\pa x_i^2} -b^2  \sum_{i< j} \frac{ 1}{(x_i-x_j)} \left( x_i^2\frac{\pa}{\pa x_i} - x_j^2\frac{\pa}{\pa x_i}  \right).
\ee
The first few Jack polynomials are:
\bea
\cC^\bet_1(x) &=& \si_1(x) \,,\non \\
\cC^\bet_2(x)&=& \si_1(x)^2 + \frac{2 }{b^2-1} \, \si_2(x)\,, \non  \\
\cC^\bet_{1^2}(x)&=& - \frac{2}{b^2 - 1} \,  \si_2(x) \,,\non \\
\cC^\bet_3(x)&=& \si_1(x)^3 + \frac{6}{ b^2 - 2} \, \si_2(x)\, \si_1(x) + \frac{6}{ (b^2-1) (b^2 - 2)}\,\si_3(x) \,, \non\\
\cC^\bet_{21}(x)&=&-\frac{6}{b^2 - 2}\,\si_2(x)\,\si_1(x) - \frac{ 18}{(2 b^2 - 1) (b^2-2)}\, \si_3(x)\,, \\
\cC^\bet_{1^3}(x)&=&\frac{ 6}{ ( b^2-1) (2b^2-1 )}\, \si_3(x) \non \,.
\eea
Here $\si_n(x) = \sum_{i_1< \cdots < x_{i_n}} x_{i_1} \cdots x_{i_n}$ are the elementary symmetric polynomials in $k$ variables. There are various different normalisations of Jack polynomials used in the literature; the normalisation above is such that $\sum_{|\xi| = d} \cC^{\bet}_{\xi}(x) = \si_1(x)^d$.

\subsection{Vertex on the strip}\label{striprules}

In this subsection we review a very useful technique for computing topological string amplitudes for geometries whose toric diagrams appear as the dual diagrams of a triangulation of a strip. This formalism was originally developed for the unrefined topological string partition functions \cite{Iqbal:2004} (to which we refer for additional details), and was later extended 
to the refined case as well \cite{Taki:2007}. This technique is particularly useful since half of the geometry giving rise to $\SU(N)$ theories with different hypermultiplets can be obtained from different triangulations of a strip (see figure \ref{sun} for an example).

Let us demonstrate the rules of the strip for the following toric diagram 

\begin{figure}[h]
\begin{center}
\begin{pspicture}(6,5)

\psline[linecolor=red,linestyle=solid,linewidth=1pt]{-}(0,1)(6,1)
\psline[linecolor=red,linestyle=solid,linewidth=1pt]{-}(0,1)(0,4)
\psline[linecolor=red,linestyle=solid,linewidth=1pt]{-}(0,4)(6,4)
\psline[linecolor=red,linestyle=solid,linewidth=1pt]{-}(6,4)(6,1)
\psline[linecolor=red,linestyle=solid,linewidth=1pt]{-}(3,1)(3,4)
\psline[linecolor=red,linestyle=solid,linewidth=1pt]{-}(0,1)(3,4)
\psline[linecolor=red,linestyle=solid,linewidth=1pt]{-}(3,4)(6,1)
\psline[linecolor=black,linestyle=solid,linewidth=2pt]{-}(-1,3)(1,3)
\psline[linecolor=black,linestyle=solid,linewidth=2pt]{-}(1,5)(1,3)
\psline[linecolor=black,linestyle=solid,linewidth=2pt]{-}(1,3)(2,2)
\psline[linecolor=black,linestyle=solid,linewidth=2pt]{-}(2,2)(4,2)
\psline[linecolor=black,linestyle=solid,linewidth=2pt]{-}(4,2)(5,3)
\psline[linecolor=black,linestyle=solid,linewidth=2pt]{-}(5,3)(7,3)
\psline[linecolor=black,linestyle=solid,linewidth=2pt]{-}(5,3)(5,5)
\psline[linecolor=black,linestyle=solid,linewidth=2pt]{-}(2,2)(2,0)
\psline[linecolor=black,linestyle=solid,linewidth=2pt]{-}(4,2)(4,0)

\put(2.1,0){$\beta_{1}$}
\put(4.1,0){$\beta_{2}$}
\put(1.1,4.7){$\alpha_{1}$}
\put(5.1,4.7){$\alpha_{2}$}
\put(1.3,2.0){$Q_{1}$}
\put(3.1,1.6){$Q_{2}$}
\put(4.7,2.4){$Q_{3}$}

\end{pspicture}
\caption{Strip} \label{charge}
\end{center}
\end{figure}

The red lines denote the triangulation of the strip and the black ones the toric diagram. The upper and lower external legs are labeled by irreducible representations $\alpha_{i}$ and $\beta_{i}$, respectively. Some of the external legs may have trivial representations, but let us keep all of them non-trivial for bookkeeping. The external legs extending in the horizontal directions are assumed to have trivial representations. We will consider the pairings of the constituent topological vertices on the strip. The contributions from the pairings depend on the local geometry. From the toric diagram, we conclude that the adjacent $\mathbb{P}^{1}$'s in the base are touching each other at one point and locally two types of line bundles over these $\mathbb{P}^{1}$'s are possible, either ${\cal O}(-1)\oplus{\cal O}(-1)\mapsto\mathbb{P}^{1}$ or ${\cal O}(-2)\oplus{\cal O}(0)\mapsto\mathbb{P}^{1}$. For each pairing, the contribution is of either of these two types.  More precisely, if in a pairing ${\cal O}(-1)\oplus{\cal O}(-1)$ appears an odd number of times the contribution is the same type, otherwise it is of type ${\cal O}(-2)\oplus{\cal O}(0)$. Note that the contributions coming from these two types of curves are each other's inverses, hence, let us write the whole amplitude in terms of one of them, say ${\cal O}(-1)\oplus{\cal O}(-1)$ and denote it by $\{\alpha\beta\}_{Q}$. For ${\cal O}(-2)\oplus{\cal O}(0)$ we then write $1/\{\alpha\beta\}_{Q}$. The subscript in this notation denotes the product of the K\"{a}hler factors involved in the pairing. The pairing is computed to be \cite{Iqbal:2004}
\bea
\{\alpha\beta\}_{Q}=\prod_{k} \left (1-Q\,q^{k} \right)^{C_{k}(\alpha,\beta)}\exp\left[\sum_{n=1}^{\infty} \frac{Q^{n}}{n(2\sin(\frac{ng_{s}}{2}))^{2}}\right],
\eea
where the exponents are defined in terms of expansion coeeficients 
\bea
\sum_k C_k(\alpha, \beta)q^k &=& \frac{q}{(q-1)^2} \left( 1 {+} (q{-}1)^2 \sum_{i=1}^{d_\alpha} q^{-i} \sum_{j=0}^{\alpha_i-1} q^j \right)\!  \left( 1 {+} (q{-}1)^2 \sum_{i=1}^{d_\beta} q^{-i} \sum_{j=0}^{\beta_i-1} q^j \right) \nonumber \\
&-&  \frac{q}{(1-q)^2}  \,.
\eea
and $d_{\alpha}$ is the number of rows of $\alpha$. Finally, we want to give the rule how to get the total amplitude using the pairings. To this end we will divide, following \cite{Iqbal:2004}, the vertices into two groups and label them by $A$ and $B$: if the first vertex on the strip has the form $C_{\mu\bullet\beta_{i}}$ (where $\mu$ belongs to an internal leg, and we label the vertex clockwise) then we call this an $A$ type vertex, otherwise, of type $B$. If two vertices are connected by a ${\mathbb P}^{1}$ with a ${\cal O}(-1)\oplus{\cal O}(-1)$ bundle over it, we change the type, otherwise not. For example, for the sequence of vertices in figure \ref{charge}, we get $(A,B,B,A)$. The total amplitude is given by all possible pairings of the external legs except the horizontal ones. Once we have associated the $A$ and $B$ labels to the vertices, the pairings are given as follows:

\bea\nonumber
i\mbox{-th vertex of type}\,A&\Longleftrightarrow& \{\beta_{i}\cdot\}\,\mbox{and}\,\{\cdot\beta_{i}^{t}\}\\ \nonumber
i\mbox{-th vertex of type}\,B&\Longleftrightarrow& \{\beta_{i}^{t}\cdot\}\,\mbox{and}\,\{\cdot\beta_{i}\}
\eea
This labeling is particularly convenient when we glue two strips to obtain a $\SU(N)$ theory. We also multiply the pairings by $s_{\beta_{i}}(q^{\rho})$ for each external leg with a non-trivial representation $\beta_{i}$.

\medskip
\medskip

\begingroup\raggedright\endgroup

\end{document}